\documentclass[twocolumn]{aastex631}
\usepackage{graphicx}
\usepackage{soul}

\shorttitle{Interior and Climate Modeling of the Venus Zone Planet TOI-2285~b}
\shortauthors{Emma L. Miles et al.}

\begin{document}

\title{Interior and Climate Modeling of the Venus Zone Planet TOI-2285~b}

\author[0009-0006-9233-1481]{Emma L. Miles}
\affiliation{Department of Earth and Planetary Sciences, University of California, Riverside, CA 92521, USA}
\email{emile006@ucr.edu}

\author[0000-0001-7968-0309]{Colby Ostberg}
\affiliation{Department of Earth and Planetary Sciences, University of California, Riverside, CA 92521, USA}

\author[0000-0002-7084-0529]{Stephen R. Kane}
\affiliation{Department of Earth and Planetary Sciences, University of California, Riverside, CA 92521, USA}

\author[0000-0003-2035-0078]{Ondrea Clarkson}
\affiliation{NASA Goddard Institute for Space Studies, 2880 Broadway, New York, NY 10025, USA}
\affiliation{Autonomic Integra, LLC, Institute for Space Studies, New York, NY 10025, USA}

\author[0000-0001-8991-3110]{Cayman T. Unterborn}
\affiliation{School of Earth and Space Exploration, Arizona State University, Tempe, AZ 85287, USA}

\author[0000-0002-3551-279X]{Tara Fetherolf}
\affiliation{Department of Earth and Planetary Sciences, University of California, Riverside, CA 92521, USA}

\author[0000-0003-3728-0475]{Michael J. Way}
\affiliation{NASA Goddard Institute for Space Studies, 2880 Broadway, New York, NY 10025, USA}
\affiliation{GSFC Sellers Exoplanet Environments Collaboration,\\ NASA Goddard Space Flight Center, MD, USA\\}
\affiliation{Theoretical Astrophysics, Department of Physics and Astronomy, Uppsala University,\\ Uppsala, SE-75120, Sweden\\}

\author[0009-0000-4164-2358]{Sadie G. Welter}
\affiliation{Department of Earth and Planetary Sciences, University of California, Riverside, CA 92521, USA}


\begin{abstract}

As the discovery of exoplanets progresses at a rapid pace, the large number of known planets provides a pathway to assess the stellar and planetary properties that govern the climate evolution of terrestrial planets. Of particular interest are those planetary cases that straddle the radius boundary of being terrestrial or gaseous in nature, such as super-Earth and sub-Neptune exoplanets, respectively. The known exoplanet, TOI-2285~b, is one such case, since it lies at the radius boundary of super-Earth and sub-Neptune ($R_p = 1.74$~$R_\oplus$), and receives a relatively high instellation flux since its orbit exists within both the Habitable Zone (HZ) and Venus Zone (VZ). Here, we present an analysis of the planetary interior and climate to determine possible evolutionary pathways for the planet. We provide volatile inventory estimates in terms of the planet's bulk density and interior composition. We performed climate simulations using ROCKE-3D that provide a suite of possible temperate scenarios for the planet for a range of topographical and initial surface water assumptions. Using the outputs of the climate simulations, we modeled JWST transmission and emission spectroscopy for each scenario. Our results demonstrate that there are temperate scenarios consistent with the known planetary properties, despite the planet's estimated steam atmosphere, and its location relative to the VZ.

\end{abstract}



\section{Introduction}
\label{intro}

 Due to improvements in ground- and space-based telescope's observational capabilities the number of detected exoplanets has drastically increased over the past few decades. Presently, more than 5500 exoplanets have been confirmed \footnote{Data accessed September 4, 2024; https://exoplanetarchive.ipac.caltech.edu/}, the majority of which have been discovered via the transit method \citep{akeson2013}. Both the Kepler Space Telescope (Kepler/K2 mission; \citep{borucki2010a,howell2014}) and the Transiting Exoplanet Survey Satellite (TESS; \cite{ricker2015}) have greatly contributed to the transiting exoplanet inventory. The transit method also favors exoplanets with shorter orbital periods, increasing the chance of detecting inner, potentially terrestrial, exoplanets \citep{kane2008b}. However, since some exoplanets are observed solely through transit observations, the analysis is often limited to knowing the radius of the planet, making it challenging to observationally distinguish between different planet types, such as super-Earths and sub-Neptune planets.

The boundary between super-Earth and sub-Neptune exoplanets is observationally distinguished by their bulk densities, which is determined from a known mass and radius measurement \citep{hu2021f}. This difference in density reflects the compositional and atmospheric differences between terrestrial and gaseous exoplanets which are shaped by distinct formation and evolutionary processes \citep{lopez2014}. However, when there is no mass measurement for a planet, this can limit the ability to classify a given planet depending on assumptions regarding either the planet mass or radius when observational data for these parameters are unavailable \citep{unterborn2016}. If the radius of a terrestrial exoplanet is measured, a mass estimate may be extracted via empirically determined mass-radius relationships \citep{weiss2014,wolfgang2016,chen2017}, with particular caution needed near the super-Earth and sub-Neptune radius boundary \citep{rogers2015a}. Therefore, other parameters must be taken into consideration to more accurately classify exoplanets at the super-Earth and sub-Neptune radius boundary, such as the quantity and state of volatile materials \citep{madhusudhan2021}. Additionally, a separation of the planetary surface conditions may be considered between those that are temperate, enabling the presence of surface liquid water, and a post-runaway greenhouse state, similar to Venus. Within the super-Earth mass regime, such surface condition dependencies should consider various factors, including atmospheric properties together with instellation flux from the host star, in order to create a complete picture of planetary evolution, as evidenced by the divergence of Venus and Earth \citep{kane2019d,gillmann2022}. Terrestrial exoplanets with shorter orbital periods may reside within the Venus Zone (VZ), defined as the radial distance from the host star in which the stellar flux may push a terrestrial exoplanet atmosphere into a runaway greenhouse state \citep{kane2014e,ostberg2019,vidaurri2022b,ostberg2023a}. Similar to the Habitable Zone (HZ), the VZ is based on an Earth-similar planet in terms of size and atmospheric properties
\citep{kasting1993a,kane2012a,kopparapu2013a,kopparapu2014,kane2016c,hill2018,hill2023}. Many planetary systems harbor super-Earths which are located within the VZ, meaning that they have the potential to display a Venus-like atmosphere. Consequently, a more accurate classification for such an exoplanet is a super-Venus \citep{kane2013d}, implying an uninhabitable rather than habitable surface environment. A detailed atmospheric characterization of super-Earth/super-Venus exoplanets, alongside other system properties, will shed light on whether super-Venus scenarios are statistically more likely than their super-Earth counterparts \citep{quirino2023}.

The recent discovery of the planet TOI-2285~b provides a natural laboratory to test various questions regarding the nature and potential surface conditions of relatively large exoplanets within the VZ \citep{fukui2022}. TOI-2285~b lies in a moderately eccentric ($e = 0.3$) orbit near the VZ/HZ boundary around an M-dwarf star, and the detection of planetary transits by TESS revealed the radius of the planet to be $\sim$1.74~$R_\oplus$, thus placing it at the radius boundary of super-Earth and sub-Neptune. Due to the relative faintness of the host star, subsequent radial velocity (RV) observations of the star were unable to recover a planetary mass measurement, but do place an upper limit of 19.5~$M_\oplus$ \citep{fukui2022}. Despite the mass ambiguity, the planet lies at multiple boundaries of potential habitability, such as planet size and instellation flux. Thus, TOI-2285~b presents an interesting case study for assessing the relative merits of classification as either a sub-Neptune or super-Earth/super-Venus.

In this paper, we present a detailed analysis of the exoplanet TOI-2285~b as an example that highlights the complications that may arise when characterizing exoplanets both near the super-Earth/sub-Neptune boundary and the overlap between the classical HZ and VZ. Section~\ref{system} describes the system properties, including the star, planet, and extent of the VZ and HZ. Section~\ref{model} presents the results of our modeling efforts for the interior and separately the climate of TOI-2285~b. The results comprise a discussion of the planetary mass, implementation of a thermodynamically self-consistent mass-radius-composition calculator, the outputs of a suite of 3D climate simulations, and predictions for spectral observations of the planet. The implications of our models for the surface nature of TOI-2285~b are discussed in Section~\ref{discussion}. Section~\ref{conclusions} summarizes the results and provides suggestions for future work.


\section{System Properties}
\label{system}


\subsection{Stellar and Planetary Parameters} 
\label{params}

TOI-2285 (TIC~329148988) is a relatively faint ($V = 13.403\pm0.092$) M-dwarf star located 42.409$\pm$0.047 away and does not currently have an estimated age. The star has a mass of 0.454~$\pm$~0.010~$M_\odot$, an effective temperature of $3491\pm58$~K, a luminosity of 0.0287$\pm$0.0010~$L_\odot$, and a metallicity, [Fe/H], of $-0.050\pm0.120$ \citep{fukui2022}. The stellar parameters of TOI-2285 can also be found in Table \ref{tab:param}. There is only one known planet orbiting TOI-2285 (referred to as TOI-2285~b). TOI-2285~b was discovered through the transit method using TESS photometry and confirmed using ground-based observatories such as SPeckle Polarimeter, TRES spectrograph, MuSCAT3 and Subaru/IRD \citep{fukui2022}. Based on the transit data, TOI-2285~b has an estimated radius of 1.74$\pm0.08$~$R_\oplus$, potentially placing the planet within the terrestrial regime. Although RV observations of the star have yet to detect the planetary signature, an upper limit for the planetary mass was placed at 19.5~$M_\oplus$.

\begin{table*}
    \centering
    \tablewidth{0.95cm}
    \caption{System Parameters of star, TOI-2285 and its planet, TOI-2285~b adapted from \citet{fukui2022}.}
    \begin{tabular}{cc}
        \textbf{Stellar Parameters}\\
        Distance to (pc) & 42.409$\pm$0.047\\ 
        Visual Magnitude & 13.403$\pm$0.092\\
        Effective Temperature (K) & 3491$\pm$58\\
        Stellar Mass ($M_\odot$) & 0.454$\pm$0.010\\
        Luminosity ($L_\odot$) & 0.0287$\pm$0.0010\\
        Metallicity (Fe/H) & -0.050$\pm$0.120\\
        \\
        \hline \\
        \textbf{Planet Parameters} \\
        Orbital Period (days) & 27.26955$^{+0.00013}_{-0.00010}$\\
        Semi-Major Axis (AU) & 0.1363$\pm$0.0010\\
        Eccentricity & 0.30$^{+0.10}_{-0.09}$ \\
        Planet Mass (M$_\oplus$) & $<$19.5 \\
        Planet Radius (R$_\oplus$) & $1.74\pm0.08$ \\
    \end{tabular}
    \label{tab:param}
\end{table*}

We consider the stellar variability of TOI-2285 in the system analysis using procedures similar to those described by \citet{fetherolf2023b,simpson2023}. We utilized the 2-minute cadence Simple Aperture Photometry (SAP) light curves that have been processed by the Science Processing Operations Center \citep[SPOC;][]{jenkins2016} and are available on the Mikulski Archive for Space Telescopes (MAST): \dataset[10.17909/t9-nmc8-f686]{https://doi:10.17909/t9-nmc8-f686}. As of the end of June 2024, the star has been observed during 7 sectors, with numerous further visits expected. The TESS data were examined using a Lomb-Scargle Periodogram \citep{lomb1976,scargle1982} and the TESS-SIP (Systematics-insensitive Periodogram) algorithm \citep{hedges2020}, from which the star was found to be generally quiet, with variations in the observed light curve being consistent with known spacecraft systematics.


\subsection{Habitable Zone and Planetary Orbit} 
\label{hz}

According to the analysis of the photometric transits and the RV measurements by \citet{fukui2022}, TOI-2285~b has an orbital period of $\sim$27.3~days, a semi-major axis of 0.136~AU, and an orbital eccentricity of $\sim$0.3 (see Figure \ref{tab:param}). It is worth noting that, since the RV observations do not detect the planetary signatures, the orbital eccentricity was determined via a fit to the photometric data, combined with an empirical upper limit on the eccentricity \citep{fukui2022}. To calculate the extent of the HZ, we use the methodology introduced by \citet{kasting1993a} and further elaborated upon by \citet{kopparapu2013a,kopparapu2014}, adopting the stellar parameters described in Section~\ref{params}. The HZ boundaries are broadly divided into the conservative region (CHZ), bounded by the runaway greenhouse and maximum greenhouse limits, and the optimistic region (OHZ), bounded by empirically-derived limits regarding the possible presence of past surface liquid water on Venus and Mars. A detailed description of how these boundaries are formulated, including the caveats and assumptions involved, are provided by \citet{kane2016c} and references therein.

\begin{figure}[hbt!]
  \includegraphics[width=8.5cm]{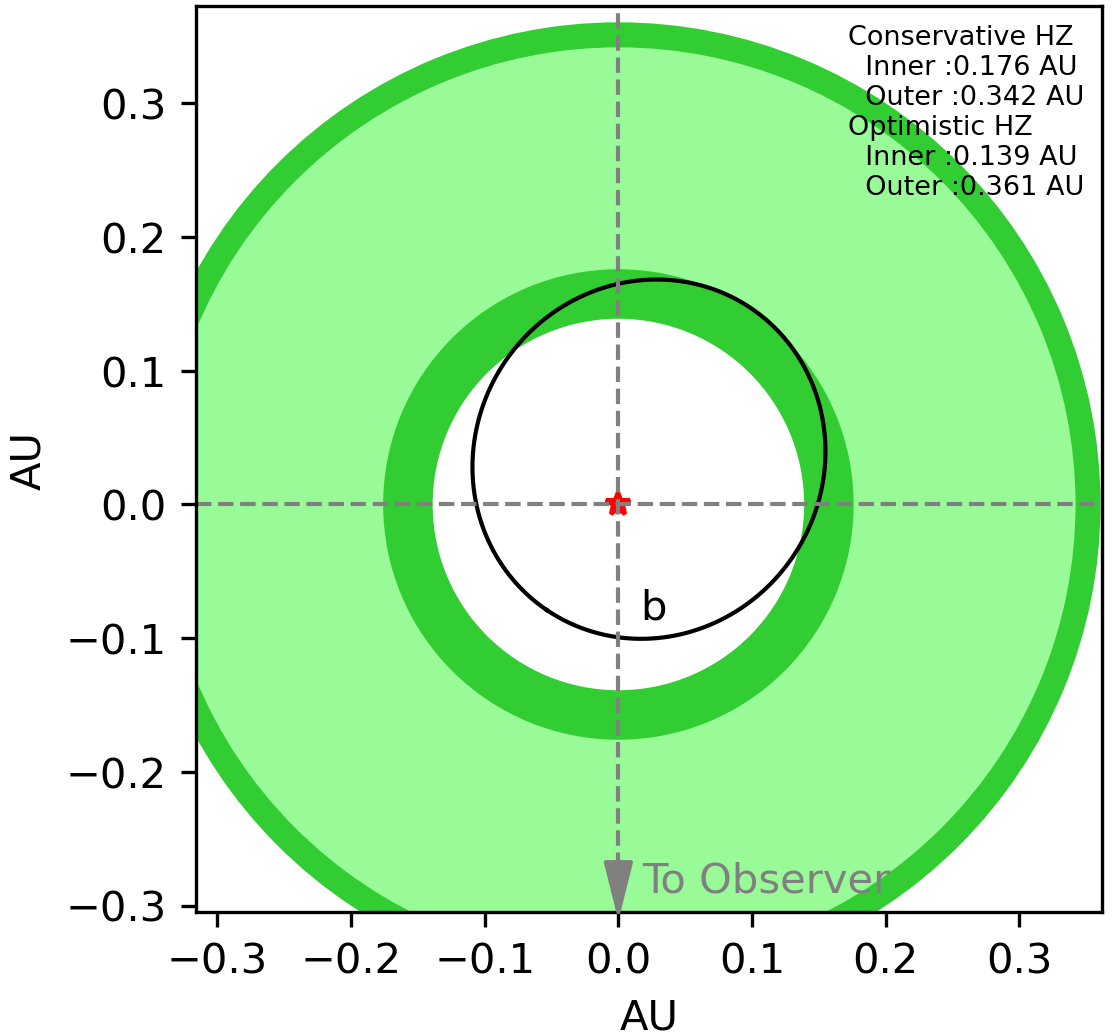}
  \caption{A top-down view of the TOI-2285 system architecture, where the scale of the figure is shown along each axis. The CHZ is shown in light green, the OHZ extensions to the HZ are shown in dark green, and the orbit of the known planet is shown as a solid line.}
  \label{fig:hz}
\end{figure}

The HZ limits of the TOI-2285 system are shown in Figure~\ref{fig:hz}, which depicts a top down view of the system. The light green region indicates the extent of the CHZ, whereas dark green shows the OHZ extensions on either side of the CHZ region. The orbit of the known planet is also shown, using the highest eccentricity ($e = 0.3$) allowed by the data described by \citet{fukui2022}, in order to demonstrate the most extreme possible orbit. This model results in \textcolor{blue}{an} apastron passage that extends into the CHZ and a periastron that occurs within the VZ, where $\sim$58\% of the orbital period lies within the OHZ. Thus, Figure~\ref{fig:hz} emphasizes that TOI-2285~b lies near the VZ/HZ boundary, and indeed may regularly cross that boundary as it orbits the host star.


\section{Modeling the Interior and Atmosphere}
\label{model}

In this section, we will describe the methodology for modeling the interior, potential climate conditions and ideal transmission and emission spectra of TOI-2285~b. In Section~\ref{mass}, we estimate the mass of TOI-2285~b using the mass-radius relationship module from Forecaster \citep{chen2017}. The following subsections show the results from interior and climate models that are used to characterize TOI-2285~b. 


\subsection{Planetary Mass Estimate}
\label{mass}

As stated in Section~\ref{system}, while the planet mass is not currently measured for TOI-2285~b, it is expected to be less than 19.5~$M_\oplus$ \citep{fukui2022}. There are various methods available to estimate the mass of an exoplanet from the measured planetary radius \citep{weiss2014,rogers2015a,wolfgang2016,unterborn2023}. Forecaster is one mass-radius relationship calculator that is used to estimate either the mass (given a known radius), or the radius (given a known mass) of an exoplanet \citep{chen2017}. This relationship is based on the distribution of measured exoplanet masses and radii which then allows the user to determine the probabilistic mass or radius, of the exoplanet of interest \citep{chen2017}. Here, we utilize Forecaster to determine a mass estimate for TOI-2285~b.

From Forecaster, the mass of TOI-2285~b is estimated to be $4.10^{+2.91}_{-1.48}$~$M_\oplus$ based on its radius. It should be noted that there is uncertainty regarding the distribution of this mass on TOI-2285~b due to the assumptions made by Forecaster. In other words, the amount of volatile inventory that the planet holds, and whether these materials are condensed is unknown. \citet{chen2017} show that the majority of planets with a radius above 1.5~$R_\oplus$ are typically classified as sub-Neptune exoplanets. This implies that above this radius threshold, all of the planet's atmospheric volatile inventory is present in non-condensed form, despite the irradiation limit for such volatile inventory \citep{turbet2020b}. This assumption limits the characterization of exoplanets given that other factors may indicate that the exoplanet may not be a gas giant, but rather a large terrestrial exoplanet like super-Earths. This means that there is a statistical bias for exoplanets at the radius boundary for super-Earth and sub-Neptune exoplanets. Given that TOI-2285~b has a radius of $1.74\pm0.08$~R$_\oplus$, this initially places TOI-2285~b in the mass-radius regime of a sub-Neptune. Upon further calculation, we find that the bulk density will be $\sim$4.28~g/cm$^3$. This density value, and the fact that TOI-2285~b lies at the outer edge of the VZ, means that it is possible TOI-2285~b is terrestrial in nature. This could imply that the Forecaster estimate is a lower limit, or that there is a high volatile inventory condensed at the surface of TOI-2285~b. Regardless of this, further investigation is required to more accurately characterize this planet.


\subsection{Interior Model} 
\label{exoplex}

ExoPlex is a Python code that calculates the mass-radius relationship, core mass fraction, bulk density, interior mineralogy and pressure-temperature profile of terrestrial exoplanets \citep{unterborn_2023}. This code caters to rocky exoplanets with radii $<2$~R$_\oplus$. It assumes a bulk molar composition, which for these models is that of the Earth's \citep{mcdonough2003}, which yields an Fe-core mass fraction of 0.33. For this work, we have updated ExoPlex to calculate the effects of a runaway steam atmosphere on a model planet's radius. We adopt the model of \citet{turbet2020a}, which calculates the height of a runaway steam atmosphere based on a planet's mass fraction of water (as steam), the radius and gravity of the rocky planet portion, the effective temperature of the atmosphere assuming it is an isothermal ideal gas, and the pressure at the transit radius. For simplicity, the Exoplex model only considers the condensation of water as a volatile, although other volatile species may condense within the planet's temperature regime \citep{unterborn_2023, Wood2019}.

Here, ExoPlex is used to calculate the bulk density of TOI-2285~b based on its planetary parameters (i.e. radius, mass and instellation flux from the host star). Given these parameters, their associated uncertainties, and the aforementioned assumptions regarding core composition, ExoPlex also statistically estimates multiple plausible atmospheric solutions. This is represented in the number of Earth oceans (1 ocean = $1.37\times10^{21}$ kg) of volatile inventory that may be present in the atmosphere. Combining this data with the calculated bulk density infers the potential volatile inventory in the atmosphere of the planet \citep{unterborn_2023}.

\begin{figure}[h!]
  \includegraphics[width=8.5cm]{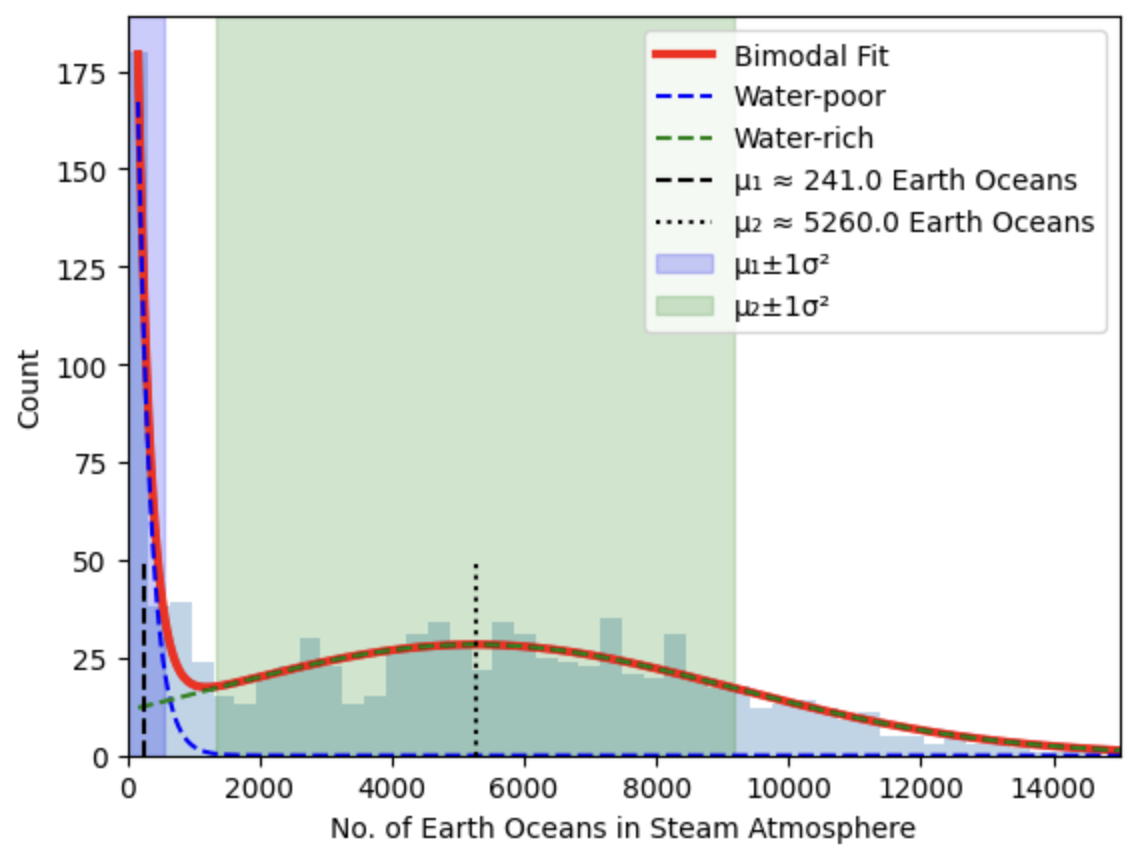}
  \caption{Histogram of number of oceans of H$_2$O as steam in a planet's atmosphere. The distribution is bimodal, indicating two distinct populations. 29\% of data points suggest TOI-2285~b is relatively dry (i.e. water-poor), with $<0.1$\% Earth oceans of water vapor in the atmosphere. A portion of the remaining 71\% correspond to a relatively water-rich atmosphere. For the water-rich population, the mean number of oceans is 5260 Earth oceans. The difference in variance between these populations further supports the distinction between the water-poor and water-rich populations.}
  \label{fig:ocean_hist}
\end{figure}

\begin{figure}[h!]
  \includegraphics[width=8.5cm]{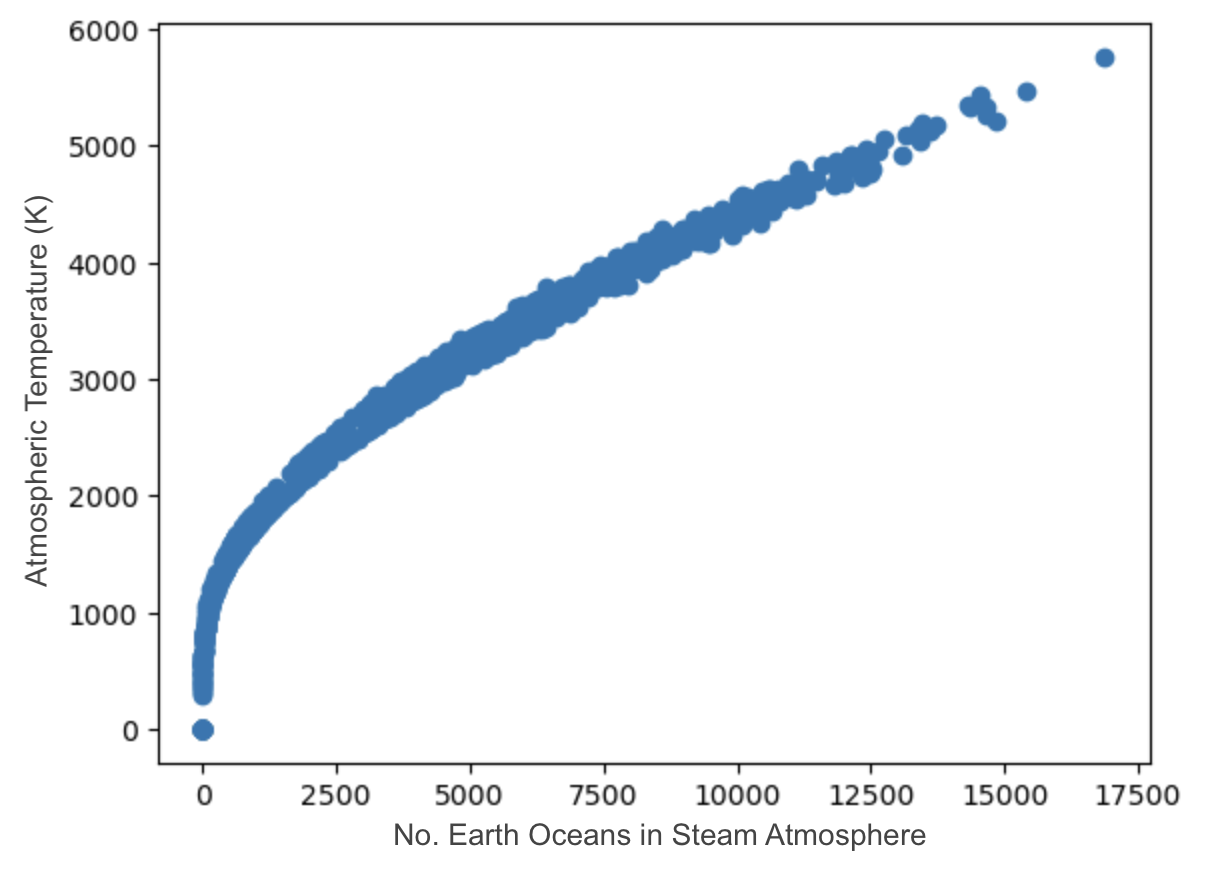}
  \caption{Relationship between effective atmospheric temperature (assuming an ideal gas)} and the number of oceans on the planet. The average number of oceans worth of steam in the atmosphere in the water-rich population was calculated to be 5260 Earth Oceans. In this plot, this equates to an atmospheric temperature of $\sim$3000~K. The average number of oceans worth of steam in the atmosphere in the extended water-poor population ($<$1000 Earth oceans) was calculated to be 241 Earth oceans (see Figure \ref{fig:ocean_hist}). In this plot, this equates to an atmospheric temperature of $\sim$1000 K.
  \label{fig:ocean_plot}
\end{figure}

Figure~\ref{fig:ocean_hist} is a histogram of the number of Earth oceans of steam in the atmosphere of TOI-2285~b. The x-axis shows the number of Earth oceans of water vapor estimated to be in the atmosphere, with values from the model grouped into bins. The y-axis indicates the number of models that fall into each bin. This distribution represents the range of interior models (and therefore, number of Earth oceans of steam in the atmosphere) that are consistent with TOI-2285~b's planetary parameters. However, the distribution of the histogram is bimodal, which implies there are two distinct populations that arise from the parameters of TOI-2285~b, where 29\% of the data points represent a water-poor population ($<0.1\%$ Earth oceans of water vapor in the atmosphere), and a portion of the remaining 71\% of the data points represent a relatively water-rich population ($>$2000 Earth oceans). To emphasize the distinction between these two populations, the mean of an extended water-poor population ($<$1000 Earth oceans), which is indicated by the spread of the Gaussian distribution for this population, is 241 Earth oceans, whereas the mean of the water-rich population is 5260 Earth oceans. The variance of each mean value was also plotted to further support the separation of these two populations. Figure \ref{fig:ocean_plot} depicts the general relationship between isothermal atmospheric temperature (assuming an ideal gas) and the number of oceans worth of steam in the atmosphere on the planet. The x axis represents the number of oceans in Earth ocean quantities, and the y axis is the temperature of the atmosphere in Kelvin (K). From Figure \ref{fig:ocean_plot}, the number of Earth oceans of steam atmosphere for TOI-2285~b determined from Figure~\ref{fig:ocean_hist} would result in an effective atmospheric temperature of $\sim3000$ K, assuming an ideal gas, which exceeds the melting point of dry peridodrite ($\sim$1300~K; \citet{katz_2003,unterborn_2023}). This makes the presence of a solid surface on TOI-2285~b unlikely. However, given the mean of the extended water-poor population, this could indicate a relatively dry atmosphere as per Figure \ref{fig:ocean_plot}.

\begin{figure}[h!]
  \includegraphics[width=8.5cm]{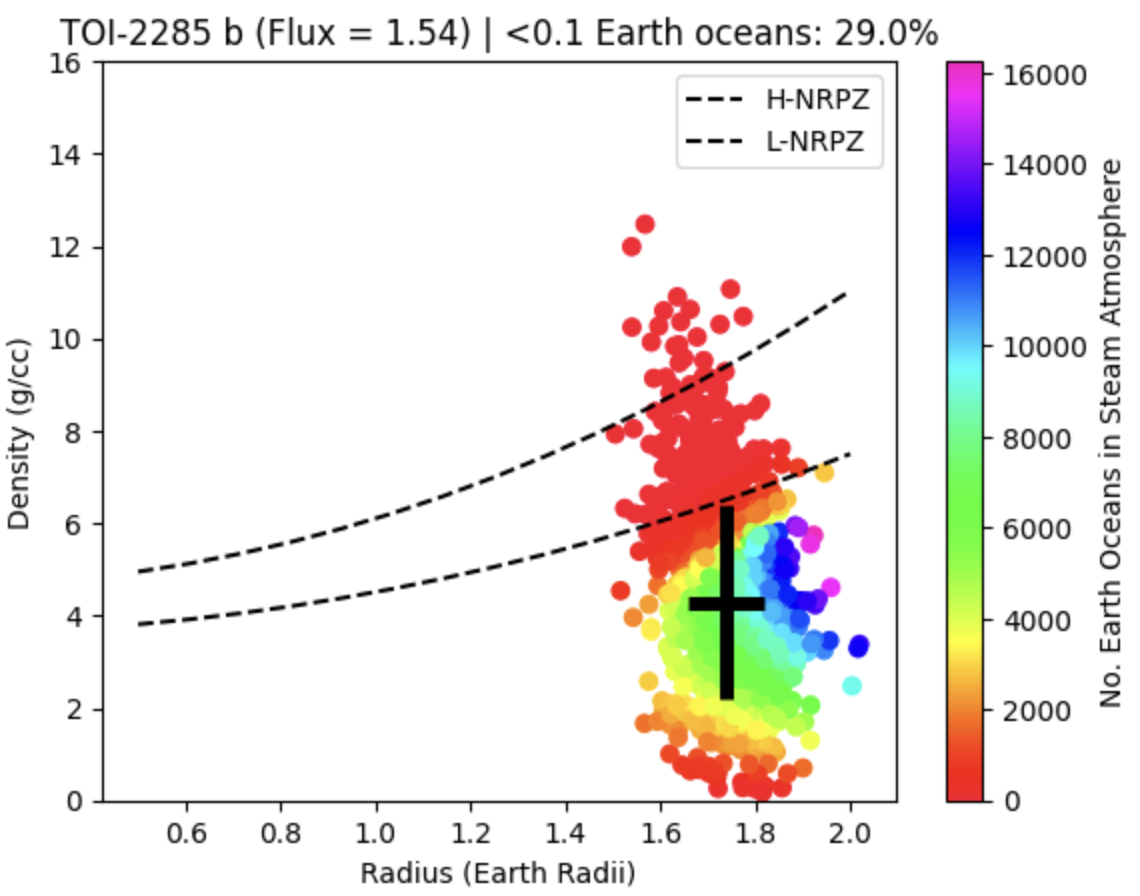}
  \caption{Plot of mass-radius relationship of TOI-2285~b with data points from Figure~\ref{fig:ocean_hist}. Color represent the number of earth oceans present as steam within the atmosphere. Dashed lines outline the nominally rocky zone of \citet{unterborn2023}, which represents the range of densities for exoplanets without significant surface volatile levels consistent with the distribution of stellar abundances. }
  \label{fig:MR_fc}
\end{figure}

Figure~\ref{fig:MR_fc} is the mass-radius relationship grid for TOI-2285~b parameters. The x-axis is the radius of the planet in Earth radii and the y-axis is the density of the planet given in g/cm$^3$. The black cross represents the limits of the density and radius uncertainties of TOI-2285~b. The colored dots represent the number of Earth oceans (i.e. volatile inventory) that the planet is likely to have in its atmosphere given its radius and density. The dotted lines are the boundary conditions known as the "Nominally Rocky Planet Zone" or NRPZ \citep{unterborn_2023}. Terrestrial exoplanets sit within the boundaries of the NRPZ. L-NRPZ is the lower boundary which separates the lowest expected stellar abundance composition from the planets with a low enough density that a volatile layer must be present. That is, it separates nominally rocky and arid planets from low density planets. H-NRPZ is the upper boundary which separates the highest expected stellar abundance composition from planets with a high enough density that it is still rocky. In other words, it separates nominally rocky planets from Super Mercury planets (large, dry terrestrial planets).

According to Figure~\ref{fig:MR_fc}, TOI-2285~b's position on the mass-radius grid sits below the L-NRPZ and H-NRPZ boundaries, indicating that it is a low density exoplanet. Due to its relatively large radius, it is expected to contain high amounts of volatile materials. This is supported by the value indicated by the color bar in Figure~\ref{fig:MR_fc} which shows that TOI-2285~b is expected to have $\sim$5000 Earth oceans of steam in its atmosphere, which also aligns with the mean value determined from \ref{fig:ocean_hist}. Therefore, TOI-2285~b may have a volatile-rich atmosphere, which is consistent with it being a sub-Neptune. However, the density error bars extend to the L-NRPZ, which crosses into the water-poor population shown in Figure \ref{fig:ocean_hist}. This implies that if TOI-2285b's mass is closer to the upper limit, resulting in a higher planet density, TOI-2285~b could be considerably dry and more terrestrial in nature. Due to the mass ambiguity, climate models are developed in Section~\ref{rocke3d} to explore the water-poor population determined from ExoPlex, and to show the potential terrestrial scenarios for TOI-2285~b.


\subsection{Planetary Climate} 
\label{rocke3d}

Here we present six 3D climate simulations run using the Resolving Orbital and Climate Keys of Earth and Extraterrestrial Environments (ROCKE-3D) code. ROCKE-3D is a three-dimensional general circulation model (GCM) developed at the NASA Goddard Institute for Space Studies \citep{way2017b} that has the capability of simulating the past and present climate conditions of terrestrial planets and moons within our Solar System, as well as predict potential climate conditions of terrestrial exoplanets given their known astrophysical parameters \citep{way2017b}. We study these simulations with the aim of exploring whether non-runaway climate states are possible on a present-day TOI-2285~b despite its location on the VZ/HZ boundary. These simulations represent a sample of possible Earth-like and (potentially) past Venusian-like climate states, as described in \citet{way2020}, with various volatile (in this case, surface water) inventories, including dynamic lakes and oceans, an aquaplanet, and a modern Earth-like configuration. We use the SOCRATES radiative transfer code \citep{1996JAtS...53.1921E,1996QJRMS.122..689E} with a synthetic stellar spectrum created by interpolating the BT-Settl grid of stellar spectra \citep{2014IAUS..299..271A,2011ASPC..448...91A}. Orbital and stellar parameters described in Section~\ref{hz} were used for all simulations. An instellation at the semi-major axis of $\mathrm{S_p} = 1.54\, \mathrm{S_\Earth}\ (2114~\mathrm{W\,m^{-2}}$) is taken from \cite{fukui2022}. As found in Section~\ref{params}, we adopt an eccentricity of 0.3, an orbital period of 27.27 days, and radius of $1.74~R_\oplus$. All simulations assume that the planet is tidally locked to its host star (e.g. in synchronous rotation). This assumption is based on the notion that planets orbiting M-dwarf stars will eventually circularize and therefore become tidally locked. However, this depends on the eccentricity and mass of the planet \citep{Barnes2017}. For the simulations presented here, we assumed that TOI-2285~b has a mass of $4.10 M_\oplus$ as estimated in Section~\ref{mass}. Higher masses become both increasingly unphysical and eventually challenging from a numerical convergence perspective, so we neglect them here in the ROCKE-3D models. 

Atmospheres were all set to 1 bar surface pressure, with the exception of one Venus Arid case, for which we tested the influence of a 10 bar atmosphere on the climate of TOI-2285~b given the same initial conditions as the 1 bar case. The composition of non-condensables is the same across all simulations and is 99.996\% N$_2$, 400~ppm CO$_2$ and 1~ppm CH$_4$. The initial topographical conditions used for the simulations broadly correspond to those found in \citet{way2020}. The Venusian topography was originally created for \cite{way2016} using Magellan mission data. Each simulation uses varying amounts of surface liquid water that fills the lowest lying points of elevation. The level of liquid water used is presented in meters (m). As stated in \citet{way2020}, the purpose of having various initial topographical conditions and levels of surface liquid water is to determine the interaction between the surface liquid water and the climate state. Venus Arid uses a modern Venus topography with no initial surface liquid water. However, there is $0.2\,\mathrm{m}$ of water embedded in the subsurface soil layers. The soil layers in this simulation are 100\% sand \citep{way2020}. The  Venus $10\,\mathrm{m}$ and Venus $310\,\mathrm{m}$ model also use modern Venus topography but contain $10\,\mathrm{m}$ and $310\,\mathrm{m}$ of initial surface liquid water. The Aqua simulation is an aquaplanet with flat bathymetry and a $158\,\mathrm{m}$ deep ocean. Finally, the Earth simulation uses a modern Earth topography with a $310\,\mathrm{m}$ deep bathtub ocean. 

\begin{deluxetable*}{|c|c|c|c|c|c|}
   \tablecolumns{11}
   \tablewidth{0pc}
   \tablecaption{\label{tab:ROCKE-3D_output} Simulations and summary of output data for TOI-2285~b using ROCKE-3D. Descriptions of the initial conditions for each simulation can be found in the text.}
   \tablehead{
     \multicolumn{1}{|c|}{Simulation ID} &
     \multicolumn{3}{|c|}{Surface Temperature ($^\circ$C)} &
     \multicolumn{1}{|c|}{$\mathrm{q_s} (\mathrm{kg\, kg^{-1}}$)} &
     \multicolumn{1}{|c|}{Planetary Albedo}}
   \startdata
    - &  Min & Max & Mean & Mean & Mean \\
    \hline
   Venus Arid &-79.6 & 79.2 & -17.5 & 3.7$\times10^{-5}$ & 26.1 \\
   Venus Arid 10 bar & -39.5 & 68.6 & 18.4 & 2.2$\times10^{-7}$ & 24.6 \\
   Venus 10m  &  -39.6 & 38.0 & -4.5 &7.8$\times10^{-4}$ & 33.6 \\
   Venus 310m & -40.7 & 22.4 & -7.9 & 4.7$\times10^{-4}$ & 34.7 \\
   Aqua  & -36.5 &  25.4 & 3.0 & 6.8$\times10^{-4}$ & 29.2 \\
   Earth  & -44.0 & 19.8 & -6.3 & 4.0$\times10^{-4}$ & 35.3 
  \enddata
\end{deluxetable*}

\begin{figure*}
     \includegraphics[width=18cm]{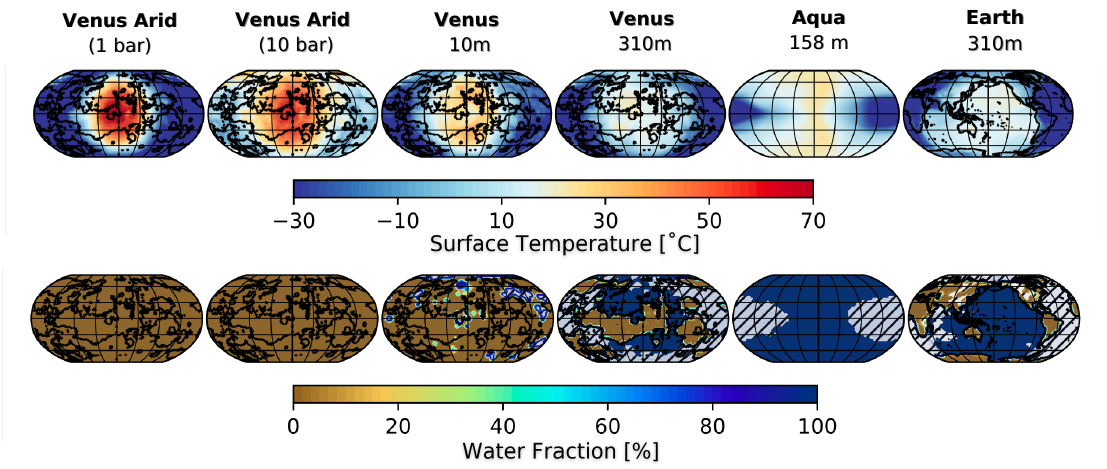}
     \caption{The ROCKE-3D GCM simulation outputs. Six climate scenarios were generated using Venus, aquaplanet and Earth topographies with varying amounts of surface water. Here, we have plotted the surface temperature, and the surface water fraction. The semi-opaque white hatched regions in the bottom row indicate regions with at least partial ice coverage.}
     \label{fig:rocke-3d_table}
\end{figure*}

All six simulations were run to the point of radiative equilibrium, which we consider to be where the top-of-the atmosphere net radiation is $\leq \pm 0.2\ \mathrm{W\,m^{-2}}$. Analysis for these simulation uses data averaged over 10 Earth years \--- nearly 134 TOI-2285~b years\--- in order to smooth out short-term variations in the climate. Table~\ref{tab:ROCKE-3D_output} and Figure~\ref{fig:rocke-3d_table} contain the output values and plots for each simulation, respectively. A well-known feature of synchronously rotating planets is substellar clouds that act to increase the planetary albedo, reflecting a large portion of the incoming stellar radiation thus stabilizing the climate \citep{2013ApJ...771L..45Y}. Such clouds form in all of our simulations to a similar extent and small differences depend on water availability.

As mentioned above, the Venus Arid simulations use a modern Venus topography with no initial liquid water on the surface. We produced two simulations using this topography: the first with a 1 bar atmosphere, and the second with a 10 bar atmosphere. This difference produces a substantial variation in the resulting climates. For the Venus Arid 1 bar simulation, there is a much higher contrast in day and night side surface temperatures as compared to the 10 bar simulation ($T_\mathrm{max}-T_\mathrm{min} \approx 159\ $and$\ 108\, ^{\circ}\mathrm{C}$, respectively (see Table~\ref{fig:rocke-3d_table}). While these precise atmospheric compositions have not been compared before, a similar study using the ExoCAM GCM \citep{2022PSJ.....3....7W} suggests that Earth-like planets in synchronous rotation around M-dwarf stars, $\mathrm{N}_{2}$-dominated atmospheres result in higher mean surface temperatures at 10 bar than those with 1 bar atmospheres, although at sub-1 bar surface pressures this trend reverses again \citep{2020ApJ...901L..36Z}. Another study done using PlaSim \citep{2005MetZe..14..299F} also reported non-monotonic temperature changes with $\mathrm{pN_2}$ and suggest that cloud feedbacks are not the likely driver for warming in their models \citep{2021Icar..35814301P}. The aforementioned studies considered fixed $\mathrm{CO_2}$ and increasing amounts of $\mathrm{N}_{2}$, resulting in a higher total pressure, whereas we increase the total pressure and keep the mixing ratios of $\mathrm{N}_{2}$ and $\mathrm{CO_2}$ fixed. Both of them had Earth like ocean-land topographies and water inventories and therefore our results are expected to differ. The warmer temperatures found in our 10 bar simulation are likely the result of greater overall greenhouse gas amounts and collisional broadening on these gases from increased pressures \citep[e.g.,][]{2009NatGe...2..891G,2020ApJ...901L..36Z, 2021Icar..35814301P}. 

The Venus $10\,\mathrm{m}$ and Venus $310\,\mathrm{m}$ simulations use a modern Venus topography with $10\,\mathrm{m}$ and $310\,\mathrm{m}$ of initial surface liquid water, respectively. These simulations are quite similar in all parameters evaluated in our study. The most notable differences are due to the surface water inventory. The Venus $310\,\mathrm{m}$ simulation has a higher surface ice fraction, shown as white hatched areas on Figure~\ref{fig:rocke-3d_table}. The night side of both planets are effectively snow covered (not shown). Despite the difference in surface water fraction shown in Figure~\ref{fig:rocke-3d_table}, both simulations have similar planetary albedo, and mean surface temperature as shown in Table \ref{tab:ROCKE-3D_output}. 

Similar to \cite{way2020}, we find the Aqua simulation has the highest mean temperatures of the 1 bar atmospheres at $3\,^{\circ}\mathrm{C}$, and simultaneously one of the lowest peak surface temperatures at the substellar point. The former is due to the effects of ocean heat transport. While a $158\,\mathrm{m}$ deep ocean does not fully capture the deep mixing that would occur on a planet with a deeper ocean (e.g. Earth), near surface heat transport creates relatively low surface temperature gradients, with a minimum and maximum temperature of $-36.5^\circ$C and $25.4^\circ$C, respectively. This is a much smaller temperature range than our other runs. Interestingly, the Aqua simulation does not have the characteristic `lobster claw' pattern on the open ocean surface as found by \cite{2014PNAS..111..629H} and \cite{delgenio2019a}. This simulation has relatively warm surface temperatures near the poles ($\sim 12\--19 ^{\circ} \mathrm{C}$) as warmer water from the substellar point is advected northward and ice covering a teardrop-like shape on the night side. We attribute the difference in open ocean morphology to the relatively shallow ocean depth compared to the aforementioned works, which allows for a warm well-mixed ice-free ocean surface covering much of the planet. 

The Earth simulation uses modern Earth topography with an ocean depth of $310\,\mathrm{m}$. \cite{delgenio2019a} showed that the artificial `eyeball' pattern commonly found in simulations using slab ocean models can be nearly recreated for a tidally locked Proxima Centauri b that has a dynamic ocean. By inhibiting ocean heat transport with an Earth land mask where the Pacific ocean is chosen as the substellar point (i.e. Asia/Australia and the Americas), it prevents warmer waters from reaching the night side of the planet. We find the same result here for our Earth simulation of TOI-2285~b. As can be seen in the bottom right-most panel in Figure~\ref{fig:rocke-3d_table}, the Pacific ocean is effectively ice free whereas the Atlantic, Arctic, and a portion of the Indian ocean are covered in ice. This further emphasizes the importance of continentality in terms of its impact on climate, particularly in the case of tidally locked worlds. This simulation has the highest mean albedo found in all of our simulations which is likely due to a combination of thick substellar clouds and substantial ice coverage.

All of our simulations fail to reach a runaway state despite the fact that TOI2285~b's orbit places it for 58$\%$ of its orbit within the OHZ\---only briefly touching the inner edge of CHZ\---and the rest of the orbit within the VZ determined by 1D models. As a cursory estimate of the onset of a moist greenhouse state, we report the mean stratospheric specific humidity, $\mathrm{q_s}$, measured at the base of the stratosphere in Table~\ref{tab:ROCKE-3D_output}. This follows work by \cite{kasting1993a}\footnote{Note that \cite{kasting1993a} used the water vapor volume mixing ratio whereas we have used the specific humidity, which under the conditions reported here should be within a factor of a few apart.}, 
although more recent models and analysis suggest that this limit should be interpreted with caution \citep[e.g.,][]{kasting2015,2019ApJ...886...16C, fujii2017b}. Many authors now report the specific humidity or mass mixing ratio at the top of the model top rather than the stratosphere. All our simulations have mean specific humidities at the model top ($~\sim0.1\,\mathrm{mb}$) of $\lesssim 7\times 10^{-6}\, \mathrm{kg\,kg^{-1}}$, the highest value occurring in the Venus 10~m run.

\subsection{Predicted Spectra} 
\label{psg}

To model the transmission and emission spectrum of each TOI-2285~b case we used the Global Emission Spectra (GlobES, \url{https://psg.gsfc.nasa.gov/apps/globes.php}), which is an application in the Planetary Spectrum Generator (PSG, \url{https://psg.gsfc.nasa.gov}) radiative transfer suite. GlobES uses the 3-D temperature-pressure (TP) and chemical abundance data from GCM simulations which allows it to incorporate the effects of an inhomogenous atmosphere and surface on both transmission and emission spectra. For each GlobES spectrum we defined the planetary, orbital, and stellar parameters to be the same as those used for the ROCKE-3D simulations see (Section~\ref{rocke3d}). 

To define the atmosphere of the six TOI-2285~b cases in GlobES we used the atmosphere data from the end of each simulation when thermal equilibrium was reached. The 3D atmosphere data that was input includes: temperature and pressure profiles; the abundances of H$_2$O, CO$_2$, CH$_4$, oxygen (O$_2$), and nitrogen (N$_2$) gases; and the abundance of liquid H$_2$O and H$_2$O-ice aerosols that were used to define the clouds. When modeling the transmission spectra, we assumed that TOI-2285~b was observed using the Near-Infrared Spectrograph (NIRSpec) PRISM instrument aboard James Webb Space Telescopes (JWST) which has a bandpass of 0.6--5.3 $\mu$m. The emission spectra assumed observations were conducted using the Mid-Infrared Instrument (MIRI) which has a bandpass of 5.0--25~$\mu$m. 

\begin{figure*}
    \centering
    \includegraphics[width=0.95\textwidth]{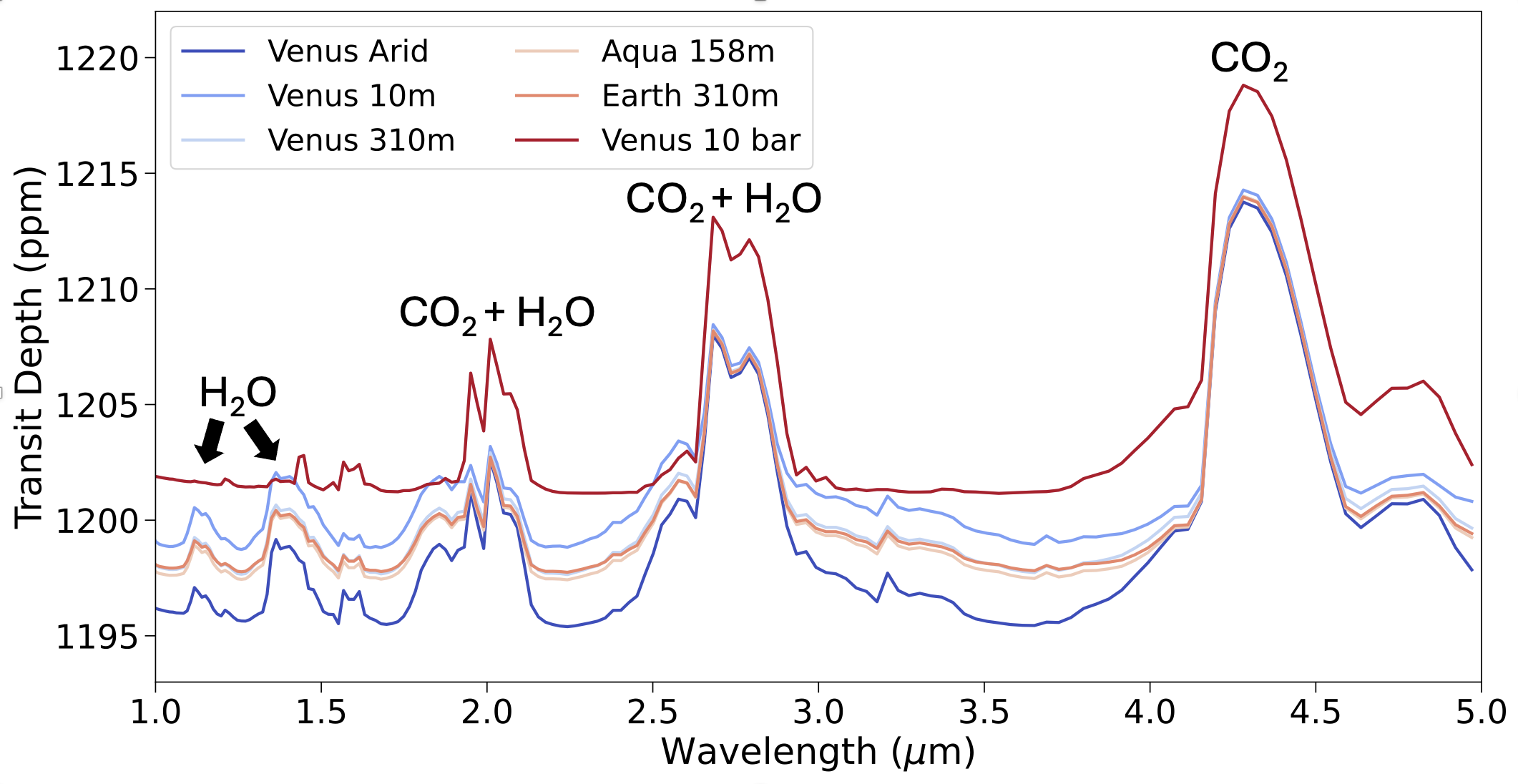}
    \caption{The GlobES modeled transmission spectra of each TOI-2285~b simulation using their final stable state atmospheres. For this model, TOI-2285~b is hypothetically observed by JWST NIRSpec PRISM. Each feature is labelled with their corresponding molecules.}
    \label{fig:TransitSpectra}
\end{figure*}

Figure~\ref{fig:TransitSpectra} shows the GlobES modelled transmission spectra for each TOI-2285~b simulation. Since the abundances of atmospheric CO$_2$, CH$_4$, and O$_2$ were defined to be the same for all six simulations, the transmission spectra all yield similar absorption features. The smaller scale height of the 10 bar atmosphere in run 10 raises the continuum of its transmission spectrum higher than the other five spectra, which conceals the smaller CO$_2$ and H$_2$O features at shorter wavelengths. The transmission spectra of the five 1-bar atmosphere simulations show some variance in the height of their continua and the size of their features.

\begin{figure*}
    \centering
    \includegraphics[width=0.9\textwidth]{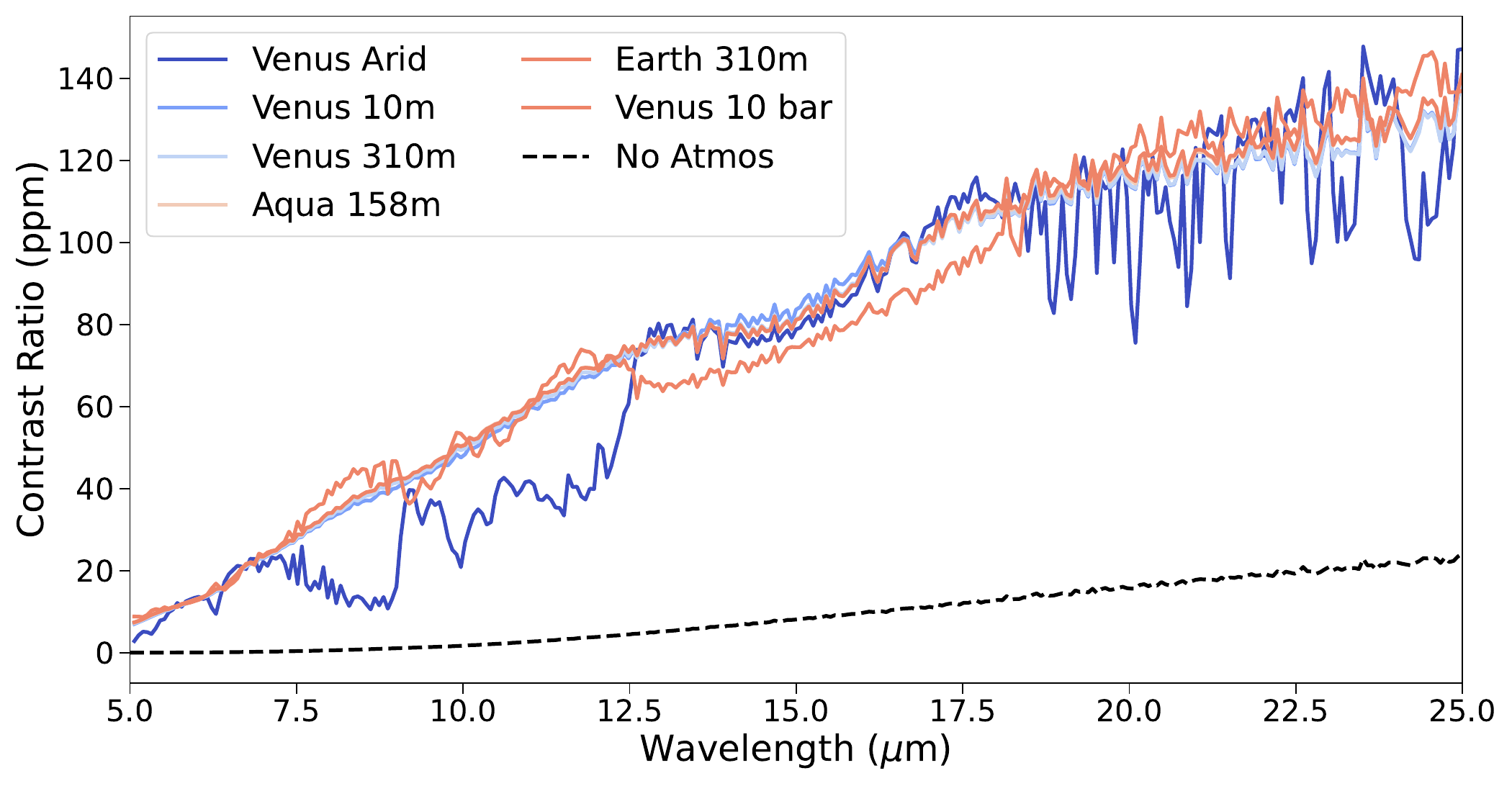}
    \caption{The GlobES modeled emission spectra of the six TOI-2285~b simulations, and a single spectrum of TOI-2285~b with no atmosphere and a T$_{eq}$ of 310 K (black line). The Venus Arid emission spectrum is the only one which has CO$_2$ features around 10 $\mu$m. The emission spectra of all cases are easily distinguished from the no atmosphere spectrum. }
    \label{fig:EmissionSpectra}
\end{figure*}

Figure~\ref{fig:EmissionSpectra} shows the modeled MIRI emission spectra for each TOI-2285~b simulation and a case where the planet has no atmosphere. There is even less variance between the emission spectrum of each simulation than there was in their transmission spectra (Figure~\ref{fig:TransitSpectra}). The Arid Venus spectrum shows CO$_2$ features around 10 $\mu$m, whereas these features are minuscule or missing from the spectra of the other simulations. The presence of features in the Arid Venus spectrum is due to the scarcity of atmospheric H$_2$O and cloud coverage compared to the dense cloud coverage in the other simulations. The no atmosphere spectrum was made assuming TOI-2285~b is a blackbody with a surface temperature of 300~K, which is its equilibrium temperature. Since the day-side temperature of the planet in the simulations were far greater than the equilibrium temperature of the planet, the resulting emission spectra are easily distinguished from the no atmosphere spectrum. This demonstrates that emissions spectroscopy with JWST may be able to efficiently determine whether TOI-2285~b has an atmosphere.

\section{Discussion}
\label{discussion}


TOI-2285~b is a transiting exoplanet with a radius of $1.74\pm0.08$~$R_\oplus$, and is located at the outer edge of the VZ of its host star. Here, we conduct a more thorough analysis of the interior and climate models produced for TOI-2285~b in Section~\ref{exoplex} and Section~\ref{rocke3d} using ExoPlex and ROCKE-3D, respecively, and further discuss the significance of this research.

\subsection{ExoPlex and ROCKE-3D} 
\label{model_discuss}

ExoPlex estimated the bulk density and the potential volatile inventory of TOI-2285~b given its known planetary parameters and assumptions of the stellar abundance of the host star. Figure~\ref{fig:MR_fc} shows that TOI-2285~b is a low density, large radii planet, and sits below the L-NRPZ and H-NRPZ boundaries. TOI-2285~b's relative position to these boundaries indicates that it may be enriched in low-density materials. Contrary to this, there is a possibility for there to be temperate surface conditions on TOI-2285~b since the density uncertainty sits at the edge of the L-NRPZ boundary. This may suggest the presence of condensed volatiles at the surface. The analysis of the ExoPlex models is dependent on the assumptions made by the calculations of this code. Since TOI-2285~b is an M-dwarf star and little is known about its composition, the stellar abundances used for the models may not accurately depict the conditions of TOI-2285~b. This aligns with the research of \citet{brinkman_2023}, where a low density planet orbiting a star that contains relatively low amounts of iron could be nominally rocky. Therefore, the possibility of terrestrial conditions on TOI-2285~b cannot be excluded.

ROCKE-3D was used to model 6 climate scenarios of TOI-2285~b. Each scenario is described in more detail in Section~\ref{rocke3d}. The models showed that a combination or radiative balance and low atmospheric water mixing ratios suggest that TOI-2285~b may not be doomed to moist or runaway greenhouse conditions. Rather than probe whether a modern Venus-like state is possible in TOI-2285~b, we constrained our work to a limited parameter space to understand how, assuming a modern Earth-like atmospheric composition,  the climate of TOI-2285~b might react to spending nearly half of its orbital period inward of the bounds of the conservative HZ. We explore the dependence of these results on initial land distribution, surface water inventory, and conduct one experiment with a 10 bar atmosphere. We have found that in all cases, we do not encounter moist greenhouse conditions, which are uninhabitable conditions resulting from high temperatures and drastic water loss \citep{kasting1988c}. Many of the models also have clement day time temperatures.

Analysis of the ExoPlex interior models in conjuction with the ROCKE-3D climate models have provided a basis for the potential surface conditions of TOI-2285~b. ExoPlex showed that TOI-2285~b may contain higher amounts of low-density materials, either on the surface or in the atmosphere. This could suggest that TOI-2285~b has an extended atmosphere or moist greenhouse conditions. However, given the density error bars in Figure~\ref{fig:MR_fc}, the planet sits at the boundary of the L-NRPZ, which could suggest a higher density, lower volatile inventory which would align with temperate conditions. In support of the potential terrestrial nature of TOI-2285~b, the climate models produced by ROCKE-3D showed that in all 6 climate scenarios, there was no indication of moist greenhouse conditions regardless of the amount of surface liquid water, and in spite of TOI-2285~b's relatively high instellation flux. Therefore, temperate conditions on TOI-2285~b are possible, but this is dependent on the initial inventory of surface liquid water on TOI-2285~b.

\subsection{PSG} 
\label{psg_discuss}

\begin{figure*}
    \centering
    \includegraphics[width=0.95\textwidth]{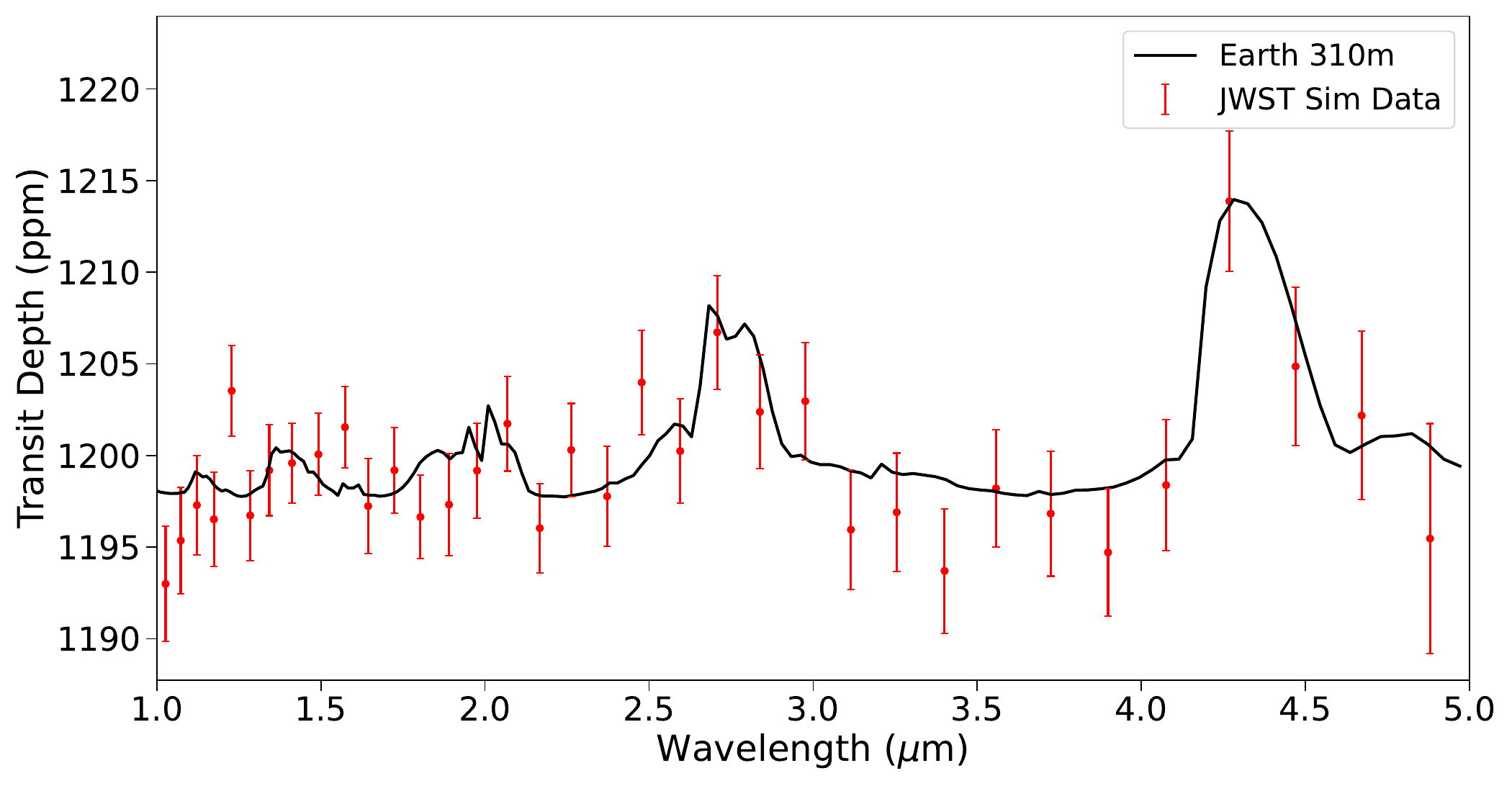}
    \caption{Simulated JWST NIRSpec PRISM data using PandExo for the Earth 310m TOI-2285~b scenario. The JWST data assumed 20 transit observations, and was binned down to a resolving power of R = 12 for easier visibility.}
    \label{fig:pandexo}
\end{figure*}

In Section~\ref{psg}, we model the transmission (Figure~\ref{fig:TransitSpectra}) and emission spectra (Figure \ref{fig:EmissionSpectra}) for each ROCKE-3D scenario using GlobES. We used PandExo \citep{batalha2017pandexo} to estimate the potential for JWST to conduct transmission spectroscopy of TOI-2285~b. PandExo uses the Space Telescope Science Institute's (STSI) Exposure Time Calculator (ETC) Pandeia to model instrumental and background noise sources of JWST observations. We defined the planet and host star in PandExo using the same parameters that were used to model the transmission and emission spectra with GlobES. The atmosphere of each TOI-2285~b scenario was defined in PandExo using their corresponding transmission spectra shown in Figure~\ref{fig:TransitSpectra}. Figure~\ref{fig:pandexo} shows simulated JWST NIRSpec PRISM data for 20 transit observations of the Earth 310m TOI-2285~b scenario. The simulated JWST data includes random scatter calculated by PandExo. The larger CO$_2$ features at 2.7 and 4.3 $\mu$m are relatively well resolved by JWST. The smaller size of the features below 2.5~$\mu$m are smaller and more difficult for JWST to resolve. 

We adopted a $\chi^2$ approach used in previous works \citep{lustigyaeger2019a,ostberg2023c} in order to calculate the signal-to-noise ratio (S/N) of the simulated JWST data for the major absorption features. We then determined the number of transits required for JWST to detect a feature with a S/N $\geq$ 5, which is the detectability threshold used in previous studies \citep[e.g.]{lustigyaeger2019a,pidhorodetska2020,ostberg2023b}. Table~\ref{tab:transits} shows the number of transit observations that JWST would require to detect 3 major absorption features in the transmission spectrum of each TOI-2285~b scenario. The other features at wavelengths less than 2.0 $\mu$m were not included in the table since they would require more than 100 transits to be detected, and are considered undetectable. The feature at 2.0~$\mu$m, which consists of CO$_2$ and H$_2$O absorption, requires extensive time to detect. It can be detected in 51 transit observations for the arid Venus scenario, and requires at least 90 transits for every other scenario. The CO$_2$ features at 2.7 and 4.3~$\mu$m are easier to detect relative to the 2.0~$\mu$m feature, as they both require between 18-31 transit observations to be detected in all scenarios. It should be noted 18 transit observations is still an extensive amount of JWST time. TOI-2285~b has a transit duration of 3.4 hours, and accounting for slew and out of transit time means 18 transit observations would require over 100 hours of JWST time.

\begin{deluxetable*}{ c c c c c c c c }
\tabletypesize{\scriptsize}
\tablecaption{Detecting TOI-2285~b Absorption Features with JWST NIRSpec PRISM}
\tablehead{\colhead{Feature} & \colhead{Wavelengths} & \multicolumn{6}{c}{Transits Required for Detection} \\
\cline{3-8}
\colhead{} & \colhead{$\mu$m} & \colhead{Venus Arid} & \colhead{Venus 10m} & \colhead{Venus 310m} & \colhead{Aqua 158m} & \colhead{Earth 310m} & \colhead{Venus 10 bar} }
\startdata
CO$_2$ + H$_2$O & 2.0 & 51 & - & 93 & - & - & 90\\
CO$_2$ + H$_2$O & 2.7 & 19 & 31 & 28 & 28 & 30 & 26\\
CO$_2$ & 4.3 & 18 & 28 & 25 & 24 & 25 & 20 \\
\enddata
\tablecomments{The number of transits that JWST NIRSpec PRISM would require to detect the major absorption features in the transmission spectrum of each TOI-2285~b scenario. The 4.3~$\mu$m CO$_2$ is the easiest feature to detect, while the CO$_2$ and H$_2$O feature at 2.0~$\mu$m would require extensive time to detect in every scenario. All the features at wavelengths less than 2.0~$\mu$m would require more than 100 transits to be detected.}
\label{tab:transits}
\end{deluxetable*}

\subsection{Modeling and Data Limitations} 
\label{caveats}

Despite the significance of the models produced for TOI-2285~b, there are several caveats that must be considered before proceeding with future research on this planet. ExoPlex is used to statistically estimate the atmospheric bulk density based on interior assumptions. In Figure~\ref{fig:ocean_hist}, we plot the weighted average which is calculated from all plausible atmospheric models for TOI-2285~b. However, the non-uniform, bimodal nature of the distribution brings a valid concern when taking the average of this data, particularly in the context of water inventory. Although the average corresponds with TOI-2285~b's position in Figure~\ref{fig:MR_fc}, it may conceal the fact that a large portion of the models calculated contain less than 1\% of atmospheric content. In future applications of ExoPlex, we plan to use a wider range of analysis to more accurately represent the range of potential interior structures.

Forecaster uses known mass-radius relationships to estimate the masses of exoplanets. In the case of TOI-2285~b, its radius correlates with the masses of sub-Neptune exoplanets, and therefore would have a density of $\sim$4.28~g/cm$^3$. This density lies between that of Mars ($\sim$3.93~g/cm$^3$) and Venus ($\sim$5.24~g/cm$^3$). However, given that the radius of Mars and Venus are significantly smaller than that of TOI-2285~b, the radius to density ratio would be greater for TOI-2285~b than for our Solar System's terrestrial planets, suggesting that TOI-2285~b may have a higher volatile inventory. The models found from Exoplex using this mass estimate also indicates that this planet may be enriched in low density materials. However, Exoplex is catered towards exoplanets orbiting solar-type stars and may not accurately depict the conditions of TOI-2285~b. Therefore, the possibility of TOI-2285~b being a terrestrial exoplanet cannot be excluded since it orbits an M-dwarf star. TOI-2285 has relatively low levels of iron (see Section~\ref{system}) compared to solar-type stars which could produce nominally rocky planets \citep{brinkman_2023}. However, the relatively large radius is likely more consistent with a non-rocky planet scenario. Thus, our ROCKE-3D simulations (see Section~\ref{rocke3d}), demonstrating possible sustained temperate climate scenarios for TOI-2285~b, address the less feasible case of a terrestrial surface, despite the calculated mass and density values.

TOI-2285~b is estimated to have an upper mass limit of 19.5~M$_\oplus$, however we simulated the climate of TOI-2285~b assuming it has a mass of $\sim$$4.10 M_\oplus$ since it becomes too numerically challenging for ROCKE-3D to simulate higher mass worlds. Masses derived using the Forecaster tool tend to have relatively large uncertainties within the terrestrial regime, and so a higher mass than that derived is consistent with the measured properties of the planet. A higher mass would lead to a more compact atmosphere with a smaller scale height than that of our simulations, but it is unclear how significantly it would affect the predicted climate. To account for this, future modeling of TOI-2285~b will implement the additional use of other 3D GCMs \citep[e.g. Laboratoire de Météorologie Dynamique Zoom (LMDZ);][]{turbet_2018} that are designed to numerically handle higher masses and surface pressures, which may produce more accurate surface conditions for large exoplanets like TOI-2285~b.

A larger mass would also decrease the scale height of the atmosphere, and would reduce the size of absorption features in its transmission spectrum. This would in turn require more JWST observation time to detect absorption features than predicted in this study (Table~\ref{tab:transits}). Additionally, we only tested TOI-2285~b with an Earth-like atmospheric composition, but it may be the case that TOI-2285~b has an atmosphere that is similar to Venus, a sub-Neptune, or even a steam-dominated hycean world. These cases were not tested since they are beyond the capability of ROCKE-3D, but should be investigated in future studies using other GCMs. A consequence of this is that the substantial steam present in the ExoPlex models were not included in the spectra generated using PSG. Including the steam in the revised GCM models, and subsequently PSG models, would likely truncate those features seen in Figure~\ref{fig:TransitSpectra} and Figure~\ref{fig:EmissionSpectra}, limiting the distinction between hazy and dessicated atmospheres \citep{lavvas2021,kempton2023a}. We also anticipate that the hazes of a Venus-like atmosphere would greatly increase the time needed for JWST to detect molecular features \citep[e.g.][]{ostberg2023c,lustigyaeger2019a,lustigyaeger2019b}, whereas a steam-dominated atmosphere without haze condensation would improve detection time \citep{harman2021snowball}. 

The radius of TOI-2285~b alone makes it difficult to infer its nature, therefore it is vital that we obtain measurements of its mass. So far, RV techniques have only been able to place upper limits on the mass of the planet due to the relative faintness of the host star. However, observations using facilities such as the NN-EXPLORE Exoplanet Investigations with Doppler Spectroscopy (NEID) spectrograph are recommended for this task since its wavelength coverage is more suitable for smaller, redder stars \citep{robertson2019}. Observations using JWST could also provide critical constraints on TOI-2285~b's atmospheric composition, and thus the specific characteristics to distinguish between a gaseous and terrestrial exoplanet. For example, identifying a H$_2$- or H$_2$O-dominated atmosphere with trace amounts of CH$_4$ and CO$_2$ might indicate that the planet has held onto its primary atmosphere, and is potentially gaseous or hycean in nature \citep{Chouqar2020, Wogan2024}. On the other hand, identifying an abundance of a combination of molecules such as water vapor, N$_2$, CO$_2$ and CH$_4$ could indicate a secondary atmospheric composition that is more terrestrial in nature \citep{Lichtenberg2025}. These distinctions are essential for understanding TOI-2285~b's formation history, and determining the type of planet that TOI-2285~b is would provide insight into the nature of planets within this radius valley \citep[e.g.][]{lopez2013role,owen2013kepler,fulton2017}.

\subsection{Significance and Future Directions} 
\label{future}

As more exoplanets with short orbital periods are discovered, we also begin to identify unique types of planets that do not exist within our Solar System. This is particularly relevant to exoplanets with a high mass and radii that sit within the VZ. The number of exoplanets being discovered in the VZ is growing substantially which is shown by the VZ catalog \citep{ostberg2023a}. Therefore, the characterization of VZ exoplanets with a higher mass and radii may reveal important information about the diversity of planets within the VZ, and enhance our understanding of planetary evolution in other systems. As stated previously, TOI-2285~b is a high radius exoplanet that sits within the VZ of its host star \citep{fukui2022}. Despite the limitations discussed in Section \ref{caveats}, the study of TOI-2285~b along with other VZ exoplanets would be useful for identifying potential trends of characteristics for VZ exoplanets. We can apply the methodology used in this paper to begin the characterization process of these exoplanets. 

The VZ catalog mentions two other exoplanets similar to TOI-2285~b with respect to having a large radius and are located in the VZ. These exoplanets are LHS~1140~c and L~98-59~d. Both exoplanets orbit M-type stars and have radii that would suggest they could be either super-Earths or sub-Neptunes \citep{cadieux2024, demangeon2021}. Unlike TOI-2285~b, both LHS~1140~c and L~98-59~d also have measured masses. Studies of LHS~1140~c suggest that this planet could be Venus-like. If it is Venus-like, detection and confirmation of transmission and spectral features for this exoplanet may be conducted in a relatively small number of transits due to its high equilibrium temperature and low surface gravity \citep{ment2019}. However, this is assuming that the exoplanet does not form a cloud or haze layer which would suppress the spectral features of atmospheric constituents like CO$_2$ \cite{barstow2020, fauchez2019, komacek2020b, ostberg_2023-exoVenus}. Further studies show that a measured eccentricity could determine if it is terrestrial or gaseous in nature \citep{gomes2020}. For L~98-59~d, there is limited analysis for this exoplanet. \citet{demangeon2021} conducted a Bayesian analysis of the internal structures of this exoplanet. They showed that this planet appears to be rich in volatile materials but whether those materials are condensed or not remains unknown \citep{demangeon2021}. Other studies emphasize that L~98-59~d sits within the VZ and could provide a benchmark for planetary characterization \citep{kostov2019b}. In both cases, there is still uncertainty regarding whether these exoplanets have terrestrial or gaseous compositions. To date, there have been no 3D climate analysis conducted on either of these planets. Considering the lack of mass constraint for TOI-2285~b limited the outcome of our models, performing a 3D climate analysis of LHS~1140~c and L~98-59~d would assist the characterization of VZ exoplanets. Given the rate at which VZ exoplanets are being identified, studying the aforementioned exoplanets further would also contribute to the future research of planetary evolution in the VZ.


\section{Conclusions}
\label{conclusions}

In this paper, we have provided a range of potential evolutionary pathways of TOI-2285~b by analyzing its planetary interior and climate using various models. In Section~\ref{mass}, a mass of $4.10^{+2.91}_{-1.48}$~$M_\oplus$ was estimated for TOI-2285~b using exoplanet mass-radius relationships since, at the present time, only an upper limit for its mass has been measured. In Section~\ref{exoplex}, we modeled the bulk density and interior composition of the planet, which show that, with the estimated planet mass, TOI-2285~b may contain higher amounts of low-density material, such as volatiles and/or an extended steam atmosphere in a runaway greenhouse state. If the planet is confirmed to have a volatile-rich atmosphere, our model predicts atmospheric temperature well in excess of the melting temperature of rock, making a solid surface layer unlikely. Roughly 30\% of interior models, however, are consistent with the planet being being terrestrial in nature, without a significant volatile envelope. As more robust mass measurements  become available, we will better be able to distinguish whether TOI-2285~b is a nominally dry or volatile-rich super-Earth that is potentially in a runaway greenhouse state. To explore the terrestrial case for TOI-2285~b, in Section~\ref{rocke3d} we presented the results of 6 climate simulations with various initial conditions and topographies. We have found that, in all cases, we do not encounter moist greenhouse conditions, and many of the models have temperate day time temperatures. In Section~\ref{psg}, we developed transmission and emission spectra for each ROCKE-3D climate scenario using PSG. Given the assumed atmospheric composition of TOI-2285~b, the transmission spectra show the large CO$_2$ features at 2.7 and 4.3 $\mu$m would be relatively well resolved if observed by JWST. The number of transits required to detect these features varies for each scenario. However, the 4.3 $\mu$m CO$_2$ feature requires a minimum of 18 transits which corresponds to over 100 hours of JWST observation time.

Although a definitive characterization of TOI-2285~b remains challenging due to insufficient information from current observations of this planet, and the relative faintness of the target, our presented methodology demonstrates the capability of present computational tools to provide first order estimates of self-consistent interior and atmospheric models. The clearest improvement to the model results would occur via further precision radial velocity measurements of the system to enable the true mass of TOI-2285~b to be extracted. Since the spectral outcome from our models are dependent on this mass, follow-up observations using JWST may yet be warranted, particularly as such observations would yield critical insight into the atmospheric state of the planet. A key aspect to assessing the possible surface conditions of the planet is its location within the VZ, which predicts a post runaway greenhouse state for a planet with a similar size and volatile inventory to Earth. If temperate conditions do indeed exist on TOI-2285~b, then the planet will become an important piece of the puzzle regarding terrestrial planet evolution. Upcoming missions to Venus, including the NASA VERITAS (Venus Emissivity, Radio Science, InSAR, Topography, and Spectroscopy) \citep{cascioli2021} and DAVINCI (Deep Atmosphere Venus Investigation of Noble gases, Chemistry, and Imaging) \citep{garvin2022} missions, and the ESA EnVision spacecraft \citep{widemann2023}, will provide a deeper understanding of the how Earth's sibling planet evolved into its present state. Thus, TOI-2285~b remains an important planet to study, and the research from this paper has provided a basis for better understanding the nature and potential surface conditions of relatively large exoplanets in the VZ. 


\section*{Acknowledgements}

The authors acknowledge support from NASA grant 80NSSC21K1797, funded through the NASA Habitable Worlds Program, and also funding support from the NASA Discovery Program for the DAVINCI Science Team. This research has made use of the Habitable Zone Gallery at hzgallery.org. The results reported herein benefited from collaborations and/or information exchange within NASA's Nexus for Exoplanet System Science (NExSS) research coordination network sponsored by NASA's Science Mission Directorate. We acknowledge the use of public TESS data from pipelines at the TESS Science Office and at the TESS Science Processing Operations Center. Resources supporting this work were provided by the NASA High-End Computing (HEC) Program through the NASA Advanced Supercomputing (NAS) Division at Ames Research Center for the production of the SPOC data products. Resources supporting this work were also provided by the NASA High-End Computing (HEC) Program through the NASA Center for Climate Simulation (NCCS) at Goddard Space Flight Center. M.J.W. acknowledges support from the GSFC Sellers Exoplanet Environments Collaboration (SEEC), which is funded by the NASA Planetary Science Division's Internal Scientist Funding Model, and ROCKE-3D which is funded by the NASA Planetary and Earth Science Divisions Internal Scientist Funding Model. All of the data presented in this paper were obtained from the Mikulski Archive for Space Telescopes (MAST). Support for MAST for non-HST data is provided by the NASA Office of Space Science via grant NNX13AC07G and by other grants and contracts. This paper includes data collected with the TESS mission, obtained from the MAST data archive at the Space Telescope Science Institute (STScI). Funding for the TESS mission is provided by the NASA Explorer Program. STScI is operated by the Association of Universities for Research in Astronomy, Inc., under NASA contract NAS 5–26555. This research made use of Lightkurve, a Python package for Kepler and TESS data analysis \citep{Lightkurve_Collaboration18}. 




\software{Mikulski Archive for Space Telescope \citep{MAST}, Lightkurve \citep{Lightkurve_Collaboration18}, Planetary Spectrum Generator \citep{Villanueva2018}, PandExo: ETC for Exoplanets \citep{batalha2017pandexo}}



\begin{thebibliography}{}
\expandafter\ifx\csname natexlab\endcsname\relax\def\natexlab#1{#1}\fi
\providecommand{\url}[1]{\href{#1}{#1}}
\providecommand{\dodoi}[1]{doi:~\href{http://doi.org/#1}{\nolinkurl{#1}}}
\providecommand{\doeprint}[1]{\href{http://ascl.net/#1}{\nolinkurl{http://ascl.net/#1}}}
\providecommand{\doarXiv}[1]{\href{https://arxiv.org/abs/#1}{\nolinkurl{https://arxiv.org/abs/#1}}}

\bibitem[{{Akeson} {et~al.}(2013){Akeson}, {Chen}, {Ciardi}, {Crane}, {Good},
  {Harbut}, {Jackson}, {Kane}, {Laity}, {Leifer}, {Lynn}, {McElroy}, {Papin},
  {Plavchan}, {Ram{\'\i}rez}, {Rey}, {von Braun}, {Wittman}, {Abajian}, {Ali},
  {Beichman}, {Beekley}, {Berriman}, {Berukoff}, {Bryden}, {Chan}, {Groom},
  {Lau}, {Payne}, {Regelson}, {Saucedo}, {Schmitz}, {Stauffer}, {Wyatt}, \&
  {Zhang}}]{akeson2013}
{Akeson}, R.~L., {Chen}, X., {Ciardi}, D., {et~al.} 2013, \pasp, 125, 989,
  \dodoi{10.1086/672273}

\bibitem[{{Allard}(2014)}]{2014IAUS..299..271A}
{Allard}, F. 2014, in Exploring the Formation and Evolution of Planetary
  Systems, ed. M.~{Booth}, B.~C. {Matthews}, \& J.~R. {Graham}, Vol. 299,
  271--272, \dodoi{10.1017/S1743921313008545}

\bibitem[{{Allard} {et~al.}(2011){Allard}, {Homeier}, \&
  {Freytag}}]{2011ASPC..448...91A}
{Allard}, F., {Homeier}, D., \& {Freytag}, B. 2011, in Astronomical Society of
  the Pacific Conference Series, Vol. 448, 16th Cambridge Workshop on Cool
  Stars, Stellar Systems, and the Sun, ed. C.~{Johns-Krull}, M.~K. {Browning},
  \& A.~A. {West}, 91.
\newblock \doarXiv{1011.5405}

\bibitem[{{Barnes}(2017)}]{Barnes2017}
{Barnes}, R. 2017, Celestial Mechanics and Dynamical Astronomy, 129, 509,
  \dodoi{10.1007/s10569-017-9783-7}

\bibitem[{{Barstow}(2020)}]{barstow2020}
{Barstow}, J.~K. 2020, \mnras, 497, 4183, \dodoi{10.1093/mnras/staa2219}

\bibitem[{Batalha {et~al.}(2017)Batalha, Mandell, Pontoppidan, Stevenson,
  Lewis, Kalirai, Earl, Greene, Albert, \& Nielsen}]{batalha2017pandexo}
Batalha, N.~E., Mandell, A., Pontoppidan, K., {et~al.} 2017, Publications of
  the Astronomical Society of the Pacific, 129, 064501

\bibitem[{{Borucki} {et~al.}(2010){Borucki}, {Koch}, {Basri}, {Batalha},
  {Brown}, {Caldwell}, {Caldwell}, {Christensen-Dalsgaard}, {Cochran},
  {DeVore}, {Dunham}, {Dupree}, {Gautier}, {Geary}, {Gilliland}, {Gould},
  {Howell}, {Jenkins}, {Kondo}, {Latham}, {Marcy}, {Meibom}, {Kjeldsen},
  {Lissauer}, {Monet}, {Morrison}, {Sasselov}, {Tarter}, {Boss}, {Brownlee},
  {Owen}, {Buzasi}, {Charbonneau}, {Doyle}, {Fortney}, {Ford}, {Holman},
  {Seager}, {Steffen}, {Welsh}, {Rowe}, {Anderson}, {Buchhave}, {Ciardi},
  {Walkowicz}, {Sherry}, {Horch}, {Isaacson}, {Everett}, {Fischer}, {Torres},
  {Johnson}, {Endl}, {MacQueen}, {Bryson}, {Dotson}, {Haas}, {Kolodziejczak},
  {Van Cleve}, {Chandrasekaran}, {Twicken}, {Quintana}, {Clarke}, {Allen},
  {Li}, {Wu}, {Tenenbaum}, {Verner}, {Bruhweiler}, {Barnes}, \&
  {Prsa}}]{borucki2010a}
{Borucki}, W.~J., {Koch}, D., {Basri}, G., {et~al.} 2010, Science, 327, 977,
  \dodoi{10.1126/science.1185402}

\bibitem[{{Brinkman} {et~al.}(2023){Brinkman}, {Weiss}, {Dai}, {Huber}, {Kite},
  {Valencia}, {Bean}, {Beard}, {Behmard}, {Blunt}, {Brady}, {Fulton},
  {Giacalone}, {Howard}, {Isaacson}, {Kasper}, {Lubin}, {MacDougall}, {Akana
  Murphy}, {Plotnykov}, {Polanski}, {Rice}, {Seifahrt}, {Stef{\'a}nsson}, \&
  {St{\"u}rmer}}]{brinkman_2023}
{Brinkman}, C.~L., {Weiss}, L.~M., {Dai}, F., {et~al.} 2023, \aj, 165, 88,
  \dodoi{10.3847/1538-3881/acad83}

\bibitem[{{Cadieux} {et~al.}(2024){Cadieux}, {Plotnykov}, {Doyon}, {Valencia},
  {Jahandar}, {Dang}, {Turbet}, {Fauchez}, {Cloutier}, {Cherubim}, {Artigau},
  {Cook}, {Edwards}, {Hallatt}, {Charnay}, {Bouchy}, {Allart}, {Mignon},
  {Baron}, {Barros}, {Benneke}, {Canto Martins}, {Cowan}, {De Medeiros},
  {Delfosse}, {Delgado-Mena}, {Dumusque}, {Ehrenreich}, {Frensch},
  {Gonz{\'a}lez Hern{\'a}ndez}, {Hara}, {Lafreni{\`e}re}, {Lo Curto}, {Malo},
  {Melo}, {Mounzer}, {Passeger}, {Pepe}, {Poulin-Girard}, {Santos},
  {Sosnowska}, {Su{\'a}rez Mascare{\~n}o}, {Thibault}, {Vaulato}, {Wade}, \&
  {Wildi}}]{cadieux2024}
{Cadieux}, C., {Plotnykov}, M., {Doyon}, R., {et~al.} 2024, \apjl, 960, L3,
  \dodoi{10.3847/2041-8213/ad1691}

\bibitem[{{Cascioli} {et~al.}(2021){Cascioli}, {Hensley}, {De Marchi},
  {Breuer}, {Durante}, {Racioppa}, {Iess}, {Mazarico}, \&
  {Smrekar}}]{cascioli2021}
{Cascioli}, G., {Hensley}, S., {De Marchi}, F., {et~al.} 2021, \psj, 2, 220,
  \dodoi{10.3847/PSJ/ac26c0}

\bibitem[{{Chen} {et~al.}(2019){Chen}, {Wolf}, {Zhan}, \&
  {Horton}}]{2019ApJ...886...16C}
{Chen}, H., {Wolf}, E.~T., {Zhan}, Z., \& {Horton}, D.~E. 2019, \apj, 886, 16,
  \dodoi{10.3847/1538-4357/ab4f7e}

\bibitem[{{Chen} \& {Kipping}(2017)}]{chen2017}
{Chen}, J., \& {Kipping}, D. 2017, \apj, 834, 17,
  \dodoi{10.3847/1538-4357/834/1/17}

\bibitem[{{Chouqar} {et~al.}(2020){Chouqar}, {Benkhaldoun}, {Jabiri},
  {Lustig-Yaeger}, {Soubkiou}, \& {Szentgyorgyi}}]{Chouqar2020}
{Chouqar}, J., {Benkhaldoun}, Z., {Jabiri}, A., {et~al.} 2020, \mnras, 495,
  962, \dodoi{10.1093/mnras/staa1198}

\bibitem[{{Del Genio} {et~al.}(2019){Del Genio}, {Way}, {Amundsen}, {Aleinov},
  {Kelley}, {Kiang}, \& {Clune}}]{delgenio2019a}
{Del Genio}, A.~D., {Way}, M.~J., {Amundsen}, D.~S., {et~al.} 2019,
  Astrobiology, 19, 99, \dodoi{10.1089/ast.2017.1760}

\bibitem[{{Demangeon} {et~al.}(2021){Demangeon}, {Zapatero Osorio}, {Alibert},
  {Barros}, {Adibekyan}, {Tabernero}, {Antoniadis-Karnavas}, {Camacho},
  {Su{\'a}rez Mascare{\~n}o}, {Oshagh}, {Micela}, {Sousa}, {Lovis}, {Pepe},
  {Rebolo}, {Cristiani}, {Santos}, {Allart}, {Allende Prieto}, {Bossini},
  {Bouchy}, {Cabral}, {Damasso}, {Di Marcantonio}, {D'Odorico}, {Ehrenreich},
  {Faria}, {Figueira}, {G{\'e}nova Santos}, {Haldemann}, {Hara}, {Gonz{\'a}lez
  Hern{\'a}ndez}, {Lavie}, {Lillo-Box}, {Lo Curto}, {Martins}, {M{\'e}gevand},
  {Mehner}, {Molaro}, {Nunes}, {Pall{\'e}}, {Pasquini}, {Poretti}, {Sozzetti},
  \& {Udry}}]{demangeon2021}
{Demangeon}, O.~D.~S., {Zapatero Osorio}, M.~R., {Alibert}, Y., {et~al.} 2021,
  \aap, 653, A41, \dodoi{10.1051/0004-6361/202140728}

\bibitem[{{Edwards}(1996)}]{1996JAtS...53.1921E}
{Edwards}, J.~M. 1996, Journal of Atmospheric Sciences, 53, 1921,
  \dodoi{10.1175/1520-0469(1996)053<1921:ECOIFA>2.0.CO;2}

\bibitem[{{Edwards} \& {Slingo}(1996)}]{1996QJRMS.122..689E}
{Edwards}, J.~M., \& {Slingo}, A. 1996, Quarterly Journal of the Royal
  Meteorological Society, 122, 689, \dodoi{10.1002/qj.49712253107}

\bibitem[{{Fauchez} {et~al.}(2019){Fauchez}, {Turbet}, {Villanueva}, {Wolf},
  {Arney}, {Kopparapu}, {Lincowski}, {Mandell}, {de Wit}, {Pidhorodetska},
  {Domagal-Goldman}, \& {Stevenson}}]{fauchez2019}
{Fauchez}, T.~J., {Turbet}, M., {Villanueva}, G.~L., {et~al.} 2019, \apj, 887,
  194, \dodoi{10.3847/1538-4357/ab5862}

\bibitem[{{Fetherolf} {et~al.}(2023){Fetherolf}, {Pepper}, {Simpson}, {Kane},
  {Mo{\v{c}}nik}, {English}, {Antoci}, {Huber}, {Jenkins}, {Stassun},
  {Twicken}, {Vanderspek}, \& {Winn}}]{fetherolf2023b}
{Fetherolf}, T., {Pepper}, J., {Simpson}, E., {et~al.} 2023, \apjs, 268, 4,
  \dodoi{10.3847/1538-4365/acdee5}

\bibitem[{{Fraedrich} {et~al.}(2005){Fraedrich}, {Jansen}, {Kirk}, {Luksch}, \&
  {Lunkeit}}]{2005MetZe..14..299F}
{Fraedrich}, K., {Jansen}, H., {Kirk}, E., {Luksch}, U., \& {Lunkeit}, F. 2005,
  Meteorologische Zeitschrift, 14, 299, \dodoi{10.1127/0941-2948/2005/0043}

\bibitem[{{Fujii} {et~al.}(2017){Fujii}, {Del Genio}, \&
  {Amundsen}}]{fujii2017b}
{Fujii}, Y., {Del Genio}, A.~D., \& {Amundsen}, D.~S. 2017, \apj, 848, 100,
  \dodoi{10.3847/1538-4357/aa8955}

\bibitem[{{Fukui} {et~al.}(2022){Fukui}, {Kimura}, {Hirano}, {Narita},
  {Kodama}, {Hori}, {Ikoma}, {Pall{\'e}}, {Murgas}, {Parviainen}, {Kawauchi},
  {Mori}, {Esparza-Borges}, {Bieryla}, {Irwin}, {Safonov}, {Stassun},
  {Alvarez-Hernandez}, {B{\'e}jar}, {Casasayas-Barris}, {Chen}, {Crouzet}, {de
  Leon}, {Isogai}, {Kagetani}, {Klagyivik}, {Korth}, {Kurita}, {Kusakabe},
  {Livingston}, {Luque}, {Madrigal-Aguado}, {Morello}, {Nishiumi},
  {Orell-Miquel}, {Oshagh}, {S{\'a}nchez-Benavente}, {Stangret}, {Terada},
  {Watanabe}, {Zou}, {Tamura}, {Kurokawa}, {Kuzuhara}, {Nishikawa}, {Omiya},
  {Vievard}, {Ueda}, {Latham}, {Quinn}, {Strakhov}, {Belinski}, {Jenkins},
  {Ricker}, {Seager}, {Vanderspek}, {Winn}, {Charbonneau}, {Ciardi}, {Collins},
  {Doty}, {Bachelet}, \& {Harbeck}}]{fukui2022}
{Fukui}, A., {Kimura}, T., {Hirano}, T., {et~al.} 2022, \pasj, 74, L1,
  \dodoi{10.1093/pasj/psab106}

\bibitem[{{Fulton} {et~al.}(2017){Fulton}, {Petigura}, {Howard}, {Isaacson},
  {Marcy}, {Cargile}, {Hebb}, {Weiss}, {Johnson}, {Morton}, {Sinukoff},
  {Crossfield}, \& {Hirsch}}]{fulton2017}
{Fulton}, B.~J., {Petigura}, E.~A., {Howard}, A.~W., {et~al.} 2017, \aj, 154,
  109, \dodoi{10.3847/1538-3881/aa80eb}

\bibitem[{{Garvin} {et~al.}(2022){Garvin}, {Getty}, {Arney}, {Johnson},
  {Kohler}, {Schwer}, {Sekerak}, {Bartels}, {Saylor}, {Elliott}, {Goodloe},
  {Garrison}, {Cottini}, {Izenberg}, {Lorenz}, {Malespin}, {Ravine}, {Webster},
  {Atkinson}, {Aslam}, {Atreya}, {Bos}, {Brinckerhoff}, {Campbell}, {Crisp},
  {Filiberto}, {Forget}, {Gilmore}, {Gorius}, {Grinspoon}, {Hofmann}, {Kane},
  {Kiefer}, {Lebonnois}, {Mahaffy}, {Pavlov}, {Trainer}, {Zahnle}, \&
  {Zolotov}}]{garvin2022}
{Garvin}, J.~B., {Getty}, S.~A., {Arney}, G.~N., {et~al.} 2022, \psj, 3, 117,
  \dodoi{10.3847/PSJ/ac63c2}

\bibitem[{{Gillmann} {et~al.}(2022){Gillmann}, {Way}, {Avice}, {Breuer},
  {Golabek}, {H{\"o}ning}, {Krissansen-Totton}, {Lammer}, {O'Rourke},
  {Persson}, {Plesa}, {Salvador}, {Scherf}, \& {Zolotov}}]{gillmann2022}
{Gillmann}, C., {Way}, M.~J., {Avice}, G., {et~al.} 2022, \ssr, 218, 56,
  \dodoi{10.1007/s11214-022-00924-0}

\bibitem[{{Goldblatt} {et~al.}(2009){Goldblatt}, {Claire}, {Lenton},
  {Matthews}, {Watson}, \& {Zahnle}}]{2009NatGe...2..891G}
{Goldblatt}, C., {Claire}, M.~W., {Lenton}, T.~M., {et~al.} 2009, Nature
  Geoscience, 2, 891, \dodoi{10.1038/ngeo692}

\bibitem[{{Gomes} \& {Ferraz-Mello}(2020)}]{gomes2020}
{Gomes}, G.~O., \& {Ferraz-Mello}, S. 2020, \mnras, 494, 5082,
  \dodoi{10.1093/mnras/staa1110}

\bibitem[{Harman {et~al.}(2021)Harman, Kopparapu, Stef{\'a}nsson, Lin,
  Mahadevan, Hedges, \& Batalha}]{harman2021snowball}
Harman, C., Kopparapu, R.~K., Stef{\'a}nsson, G., {et~al.} 2021, arXiv preprint
  arXiv:2109.10838

\bibitem[{{Hedges} {et~al.}(2020){Hedges}, {Angus}, {Barentsen}, {Saunders},
  {Montet}, \& {Gully-Santiago}}]{hedges2020}
{Hedges}, C., {Angus}, R., {Barentsen}, G., {et~al.} 2020, Research Notes of
  the American Astronomical Society, 4, 220, \dodoi{10.3847/2515-5172/abd106}

\bibitem[{{Hill} {et~al.}(2023){Hill}, {Bott}, {Dalba}, {Fetherolf}, {Kane},
  {Kopparapu}, {Li}, \& {Ostberg}}]{hill2023}
{Hill}, M.~L., {Bott}, K., {Dalba}, P.~A., {et~al.} 2023, \aj, 165, 34,
  \dodoi{10.3847/1538-3881/aca1c0}

\bibitem[{{Hill} {et~al.}(2018){Hill}, {Kane}, {Seperuelo Duarte}, {Kopparapu},
  {Gelino}, \& {Wittenmyer}}]{hill2018}
{Hill}, M.~L., {Kane}, S.~R., {Seperuelo Duarte}, E., {et~al.} 2018, \apj, 860,
  67, \dodoi{10.3847/1538-4357/aac384}

\bibitem[{{Howell} {et~al.}(2014){Howell}, {Sobeck}, {Haas}, {Still},
  {Barclay}, {Mullally}, {Troeltzsch}, {Aigrain}, {Bryson}, {Caldwell},
  {Chaplin}, {Cochran}, {Huber}, {Marcy}, {Miglio}, {Najita}, {Smith},
  {Twicken}, \& {Fortney}}]{howell2014}
{Howell}, S.~B., {Sobeck}, C., {Haas}, M., {et~al.} 2014, \pasp, 126, 398,
  \dodoi{10.1086/676406}

\bibitem[{{Hu} {et~al.}(2021){Hu}, {Damiano}, {Scheucher}, {Kite}, {Seager}, \&
  {Rauer}}]{hu2021f}
{Hu}, R., {Damiano}, M., {Scheucher}, M., {et~al.} 2021, \apjl, 921, L8,
  \dodoi{10.3847/2041-8213/ac1f92}

\bibitem[{{Hu} \& {Yang}(2014)}]{2014PNAS..111..629H}
{Hu}, Y., \& {Yang}, J. 2014, Proceedings of the National Academy of Science,
  111, 629, \dodoi{10.1073/pnas.1315215111}

\bibitem[{{Jenkins} {et~al.}(2016){Jenkins}, {Twicken}, {McCauliff},
  {Campbell}, {Sanderfer}, {Lung}, {Mansouri-Samani}, {Girouard}, {Tenenbaum},
  {Klaus}, {Smith}, {Caldwell}, {Chacon}, {Henze}, {Heiges}, {Latham},
  {Morgan}, {Swade}, {Rinehart}, \& {Vanderspek}}]{jenkins2016}
{Jenkins}, J.~M., {Twicken}, J.~D., {McCauliff}, S., {et~al.} 2016, in
  \procspie, Vol. 9913, Software and Cyberinfrastructure for Astronomy IV,
  99133E, \dodoi{10.1117/12.2233418}

\bibitem[{{Kane} {et~al.}(2013){Kane}, {Barclay}, \& {Gelino}}]{kane2013d}
{Kane}, S.~R., {Barclay}, T., \& {Gelino}, D.~M. 2013, \apjl, 770, L20,
  \dodoi{10.1088/2041-8205/770/2/L20}

\bibitem[{{Kane} \& {Gelino}(2012)}]{kane2012a}
{Kane}, S.~R., \& {Gelino}, D.~M. 2012, \pasp, 124, 323, \dodoi{10.1086/665271}

\bibitem[{{Kane} {et~al.}(2014){Kane}, {Kopparapu}, \&
  {Domagal-Goldman}}]{kane2014e}
{Kane}, S.~R., {Kopparapu}, R.~K., \& {Domagal-Goldman}, S.~D. 2014, \apjl,
  794, L5, \dodoi{10.1088/2041-8205/794/1/L5}

\bibitem[{{Kane} \& {von Braun}(2008)}]{kane2008b}
{Kane}, S.~R., \& {von Braun}, K. 2008, \apj, 689, 492, \dodoi{10.1086/592381}

\bibitem[{{Kane} {et~al.}(2016){Kane}, {Hill}, {Kasting}, {Kopparapu},
  {Quintana}, {Barclay}, {Batalha}, {Borucki}, {Ciardi}, {Haghighipour},
  {Hinkel}, {Kaltenegger}, {Selsis}, \& {Torres}}]{kane2016c}
{Kane}, S.~R., {Hill}, M.~L., {Kasting}, J.~F., {et~al.} 2016, \apj, 830, 1,
  \dodoi{10.3847/0004-637X/830/1/1}

\bibitem[{{Kane} {et~al.}(2019){Kane}, {Arney}, {Crisp}, {Domagal-Goldman},
  {Glaze}, {Goldblatt}, {Grinspoon}, {Head}, {Lenardic}, {Unterborn}, {Way}, \&
  {Zahnle}}]{kane2019d}
{Kane}, S.~R., {Arney}, G., {Crisp}, D., {et~al.} 2019, Journal of Geophysical
  Research (Planets), 124, 2015, \dodoi{10.1029/2019JE005939}

\bibitem[{{Kasting}(1988)}]{kasting1988c}
{Kasting}, J.~F. 1988, \icarus, 74, 472, \dodoi{10.1016/0019-1035(88)90116-9}

\bibitem[{{Kasting} {et~al.}(2015){Kasting}, {Chen}, \&
  {Kopparapu}}]{kasting2015}
{Kasting}, J.~F., {Chen}, H., \& {Kopparapu}, R.~K. 2015, \apjl, 813, L3,
  \dodoi{10.1088/2041-8205/813/1/L3}

\bibitem[{{Kasting} {et~al.}(1993){Kasting}, {Whitmire}, \&
  {Reynolds}}]{kasting1993a}
{Kasting}, J.~F., {Whitmire}, D.~P., \& {Reynolds}, R.~T. 1993, \icarus, 101,
  108, \dodoi{10.1006/icar.1993.1010}

\bibitem[{{Katz} {et~al.}(2003){Katz}, {Spiegelman}, \& {Langmuir}}]{katz_2003}
{Katz}, R.~F., {Spiegelman}, M., \& {Langmuir}, C.~H. 2003, Geochemistry,
  Geophysics, Geosystems, 4, 1073, \dodoi{10.1029/2002GC000433}

\bibitem[{{Kempton} {et~al.}(2023){Kempton}, {Lessard}, {Malik}, {Rogers},
  {Futrowsky}, {Ih}, {Marounina}, \& {Mu{\~n}oz-Romero}}]{kempton2023a}
{Kempton}, E. M.~R., {Lessard}, M., {Malik}, M., {et~al.} 2023, \apj, 953, 57,
  \dodoi{10.3847/1538-4357/ace10d}

\bibitem[{{Komacek} {et~al.}(2020){Komacek}, {Fauchez}, {Wolf}, \&
  {Abbot}}]{komacek2020b}
{Komacek}, T.~D., {Fauchez}, T.~J., {Wolf}, E.~T., \& {Abbot}, D.~S. 2020,
  \apjl, 888, L20, \dodoi{10.3847/2041-8213/ab6200}

\bibitem[{{Kopparapu} {et~al.}(2014){Kopparapu}, {Ramirez}, {SchottelKotte},
  {Kasting}, {Domagal-Goldman}, \& {Eymet}}]{kopparapu2014}
{Kopparapu}, R.~K., {Ramirez}, R.~M., {SchottelKotte}, J., {et~al.} 2014, \apj,
  787, L29, \dodoi{10.1088/2041-8205/787/2/L29}

\bibitem[{{Kopparapu} {et~al.}(2013){Kopparapu}, {Ramirez}, {Kasting}, {Eymet},
  {Robinson}, {Mahadevan}, {Terrien}, {Domagal-Goldman}, {Meadows}, \&
  {Deshpande}}]{kopparapu2013a}
{Kopparapu}, R.~K., {Ramirez}, R., {Kasting}, J.~F., {et~al.} 2013, \apj, 765,
  131, \dodoi{10.1088/0004-637X/765/2/131}

\bibitem[{{Kostov} {et~al.}(2019){Kostov}, {Schlieder}, {Barclay}, {Quintana},
  {Col{\'o}n}, {Brande}, {Collins}, {Feinstein}, {Hadden}, {Kane}, {Kreidberg},
  {Kruse}, {Lam}, {Matthews}, {Montet}, {Pozuelos}, {Stassun}, {Winters},
  {Ricker}, {Vanderspek}, {Latham}, {Seager}, {Winn}, {Jenkins}, {Afanasev},
  {Armstrong}, {Arney}, {Boyd}, {Barentsen}, {Barkaoui}, {Batalha}, {Beichman},
  {Bayliss}, {Burke}, {Burdanov}, {Cacciapuoti}, {Carson}, {Charbonneau},
  {Christiansen}, {Ciardi}, {Clampin}, {Collins}, {Conti}, {Coughlin},
  {Covone}, {Crossfield}, {Delrez}, {Domagal-Goldman}, {Dressing}, {Ducrot},
  {Essack}, {Everett}, {Fauchez}, {Foreman-Mackey}, {Gan}, {Gilbert}, {Gillon},
  {Gonzales}, {Hamann}, {Hedges}, {Hocutt}, {Hoffman}, {Horch}, {Horne},
  {Howell}, {Hynes}, {Ireland}, {Irwin}, {Isopi}, {Jensen}, {Jehin},
  {Kaltenegger}, {Kielkopf}, {Kopparapu}, {Lewis}, {Lopez}, {Lissauer}, {Mann},
  {Mallia}, {Mandell}, {Matson}, {Mazeh}, {Monsue}, {Moran}, {Moran}, {Morley},
  {Morris}, {Muirhead}, {Mukai}, {Mullally}, {Mullally}, {Murray}, {Narita},
  {Palle}, {Pidhorodetska}, {Quinn}, {Relles}, {Rinehart}, {Ritsko},
  {Rodriguez}, {Rowden}, {Rowe}, {Sebastian}, {Sefako}, {Shahaf}, {Shporer},
  {Ta{\~n}{\'o}n Reyes}, {Tenenbaum}, {Ting}, {Twicken}, {van Belle}, {Vega},
  {Volosin}, {Walkowicz}, \& {Youngblood}}]{kostov2019b}
{Kostov}, V.~B., {Schlieder}, J.~E., {Barclay}, T., {et~al.} 2019, \aj, 158,
  32, \dodoi{10.3847/1538-3881/ab2459}

\bibitem[{{Lavvas} \& {Arfaux}(2021)}]{lavvas2021}
{Lavvas}, P., \& {Arfaux}, A. 2021, \mnras, 502, 5643,
  \dodoi{10.1093/mnras/stab456}

\bibitem[{{Lichtenberg} \& {Miguel}(2025)}]{Lichtenberg2025}
{Lichtenberg}, T., \& {Miguel}, Y. 2025, Treatise on Geochemistry, 7, 51,
  \dodoi{10.1016/B978-0-323-99762-1.00122-4}

\bibitem[{{Lightkurve Collaboration} {et~al.}(2018){Lightkurve Collaboration},
  {Cardoso}, {Hedges}, {Gully-Santiago}, {Saunders}, {Cody}, {Barclay}, {Hall},
  {Sagear}, {Turtelboom}, {Zhang}, {Tzanidakis}, {Mighell}, {Coughlin}, {Bell},
  {Berta-Thompson}, {Williams}, {Dotson}, \&
  {Barentsen}}]{Lightkurve_Collaboration18}
{Lightkurve Collaboration}, {Cardoso}, J.~V.~d.~M., {Hedges}, C., {et~al.}
  2018, {Lightkurve: Kepler and TESS time series analysis in Python},
  Astrophysics Source Code Library.
\newblock \doeprint{1812.013}

\bibitem[{{Lomb}(1976)}]{lomb1976}
{Lomb}, N.~R. 1976, \apss, 39, 447, \dodoi{10.1007/BF00648343}

\bibitem[{Lopez \& Fortney(2013)}]{lopez2013role}
Lopez, E.~D., \& Fortney, J.~J. 2013, The Astrophysical Journal, 776, 2

\bibitem[{{Lopez} \& {Fortney}(2014)}]{lopez2014}
{Lopez}, E.~D., \& {Fortney}, J.~J. 2014, \apj, 792, 1,
  \dodoi{10.1088/0004-637X/792/1/1}

\bibitem[{{Lustig-Yaeger} {et~al.}(2019{\natexlab{a}}){Lustig-Yaeger},
  {Meadows}, \& {Lincowski}}]{lustigyaeger2019a}
{Lustig-Yaeger}, J., {Meadows}, V.~S., \& {Lincowski}, A.~P.
  2019{\natexlab{a}}, \aj, 158, 27, \dodoi{10.3847/1538-3881/ab21e0}

\bibitem[{{Lustig-Yaeger} {et~al.}(2019{\natexlab{b}}){Lustig-Yaeger},
  {Meadows}, \& {Lincowski}}]{lustigyaeger2019b}
---. 2019{\natexlab{b}}, \apjl, 887, L11, \dodoi{10.3847/2041-8213/ab5965}

\bibitem[{{Madhusudhan} {et~al.}(2021){Madhusudhan}, {Piette}, \&
  {Constantinou}}]{madhusudhan2021}
{Madhusudhan}, N., {Piette}, A. A.~A., \& {Constantinou}, S. 2021, \apj, 918,
  1, \dodoi{10.3847/1538-4357/abfd9c}

\bibitem[{{McDonough}(2003)}]{mcdonough2003}
{McDonough}, W.~F. 2003, Treatise on Geochemistry, 2, 568,
  \dodoi{10.1016/B0-08-043751-6/02015-6}

\bibitem[{{Ment} {et~al.}(2019){Ment}, {Dittmann}, {Astudillo-Defru},
  {Charbonneau}, {Irwin}, {Bonfils}, {Murgas}, {Almenara}, {Forveille}, {Agol},
  {Ballard}, {Berta-Thompson}, {Bouchy}, {Cloutier}, {Delfosse}, {Doyon},
  {Dressing}, {Esquerdo}, {Haywood}, {Kipping}, {Latham}, {Lovis}, {Newton},
  {Pepe}, {Rodriguez}, {Santos}, {Tan}, {Udry}, {Winters}, \&
  {W{\"u}nsche}}]{ment2019}
{Ment}, K., {Dittmann}, J.~A., {Astudillo-Defru}, N., {et~al.} 2019, \aj, 157,
  32, \dodoi{10.3847/1538-3881/aaf1b1}

\bibitem[{{Ostberg} \& {Kane}(2019)}]{ostberg2019}
{Ostberg}, C., \& {Kane}, S.~R. 2019, \aj, 158, 195,
  \dodoi{10.3847/1538-3881/ab44b0}

\bibitem[{{Ostberg} {et~al.}(2023{\natexlab{a}}){Ostberg}, {Kane}, {Lincowski},
  \& {Dalba}}]{ostberg2023c}
{Ostberg}, C., {Kane}, S.~R., {Lincowski}, A.~P., \& {Dalba}, P.~A.
  2023{\natexlab{a}}, \aj, 166, 213, \dodoi{10.3847/1538-3881/acfed2}

\bibitem[{{Ostberg} {et~al.}(2023{\natexlab{b}}){Ostberg}, {Kane}, {Lincowski},
  \& {Dalba}}]{ostberg_2023-exoVenus}
---. 2023{\natexlab{b}}, \aj, 166, 213, \dodoi{10.3847/1538-3881/acfed2}

\bibitem[{{Ostberg} {et~al.}(2023{\natexlab{c}}){Ostberg}, {Kane}, {Li},
  {Schwieterman}, {Hill}, {Bott}, {Dalba}, {Fetherolf}, {Head}, \&
  {Unterborn}}]{ostberg2023a}
{Ostberg}, C., {Kane}, S.~R., {Li}, Z., {et~al.} 2023{\natexlab{c}}, \aj, 165,
  168, \dodoi{10.3847/1538-3881/acbfaf}

\bibitem[{{Ostberg} {et~al.}(2023{\natexlab{d}}){Ostberg}, {Guzewich}, {Kane},
  {Kohler}, {Oman}, {Fauchez}, {Kopparapu}, {Richardson}, \&
  {Whelley}}]{ostberg2023b}
{Ostberg}, C.~M., {Guzewich}, S.~D., {Kane}, S.~R., {et~al.}
  2023{\natexlab{d}}, \aj, 166, 199, \dodoi{10.3847/1538-3881/acfe12}

\bibitem[{Owen \& Wu(2013)}]{owen2013kepler}
Owen, J.~E., \& Wu, Y. 2013, The Astrophysical Journal, 775, 105

\bibitem[{{Paradise} {et~al.}(2021){Paradise}, {Fan}, {Menou}, \&
  {Lee}}]{2021Icar..35814301P}
{Paradise}, A., {Fan}, B.~L., {Menou}, K., \& {Lee}, C. 2021, \icarus, 358,
  114301, \dodoi{10.1016/j.icarus.2020.114301}

\bibitem[{{Pidhorodetska} {et~al.}(2020){Pidhorodetska}, {Fauchez},
  {Villanueva}, {Domagal-Goldman}, \& {Kopparapu}}]{pidhorodetska2020}
{Pidhorodetska}, D., {Fauchez}, T.~J., {Villanueva}, G.~L., {Domagal-Goldman},
  S.~D., \& {Kopparapu}, R.~K. 2020, \apjl, 898, L33,
  \dodoi{10.3847/2041-8213/aba4a1}

\bibitem[{{Quirino} {et~al.}(2023){Quirino}, {Gilli}, {Kaltenegger}, {Navarro},
  {Fauchez}, {Turbet}, {Leconte}, {Lebonnois}, \&
  {Gonz{\'a}lez-Galindo}}]{quirino2023}
{Quirino}, D., {Gilli}, G., {Kaltenegger}, L., {et~al.} 2023, \mnras, 523, L86,
  \dodoi{10.1093/mnrasl/slad045}

\bibitem[{{Ricker} {et~al.}(2015){Ricker}, {Winn}, {Vanderspek}, {Latham},
  {Bakos}, {Bean}, {Berta-Thompson}, {Brown}, {Buchhave}, {Butler}, {Butler},
  {Chaplin}, {Charbonneau}, {Christensen-Dalsgaard}, {Clampin}, {Deming},
  {Doty}, {De Lee}, {Dressing}, {Dunham}, {Endl}, {Fressin}, {Ge}, {Henning},
  {Holman}, {Howard}, {Ida}, {Jenkins}, {Jernigan}, {Johnson}, {Kaltenegger},
  {Kawai}, {Kjeldsen}, {Laughlin}, {Levine}, {Lin}, {Lissauer}, {MacQueen},
  {Marcy}, {McCullough}, {Morton}, {Narita}, {Paegert}, {Palle}, {Pepe},
  {Pepper}, {Quirrenbach}, {Rinehart}, {Sasselov}, {Sato}, {Seager},
  {Sozzetti}, {Stassun}, {Sullivan}, {Szentgyorgyi}, {Torres}, {Udry}, \&
  {Villasenor}}]{ricker2015}
{Ricker}, G.~R., {Winn}, J.~N., {Vanderspek}, R., {et~al.} 2015, Journal of
  Astronomical Telescopes, Instruments, and Systems, 1, 014003,
  \dodoi{10.1117/1.JATIS.1.1.014003}

\bibitem[{{Robertson} {et~al.}(2019){Robertson}, {Anderson}, {Stefansson},
  {Hearty}, {Monson}, {Mahadevan}, {Blakeslee}, {Bender}, {Ninan}, {Conran},
  {Levi}, {Lubar}, {Cole}, {Dykhouse}, {Kanodia}, {Nitroy}, {Smolsky},
  {Tuggle}, {Blank}, {Nelson}, {Blake}, {Halverson}, {Henderson}, {Kaplan},
  {Li}, {Logsdon}, {McElwain}, {Rajagopal}, {Ramsey}, {Roy}, {Schwab},
  {Terrien}, \& {Wright}}]{robertson2019}
{Robertson}, P., {Anderson}, T., {Stefansson}, G., {et~al.} 2019, Journal of
  Astronomical Telescopes, Instruments, and Systems, 5, 015003,
  \dodoi{10.1117/1.JATIS.5.1.015003}

\bibitem[{{Rogers}(2015)}]{rogers2015a}
{Rogers}, L.~A. 2015, \apj, 801, 41, \dodoi{10.1088/0004-637X/801/1/41}

\bibitem[{{Scargle}(1982)}]{scargle1982}
{Scargle}, J.~D. 1982, \apj, 263, 835, \dodoi{10.1086/160554}

\bibitem[{{Simpson} {et~al.}(2023){Simpson}, {Fetherolf}, {Kane}, {Pepper},
  {Mo{\v{c}}nik}, \& {Dalba}}]{simpson2023}
{Simpson}, E.~R., {Fetherolf}, T., {Kane}, S.~R., {et~al.} 2023, \aj, 166, 72,
  \dodoi{10.3847/1538-3881/acda26}

\bibitem[{Team(2021)}]{MAST}
Team, M. 2021, TESS Light Curves - All Sectors,  STScI/MAST,
  \dodoi{10.17909/T9-NMC8-F686}

\bibitem[{{Turbet} {et~al.}(2020{\natexlab{a}}){Turbet}, {Bolmont},
  {Ehrenreich}, {Gratier}, {Leconte}, {Selsis}, {Hara}, \&
  {Lovis}}]{turbet2020b}
{Turbet}, M., {Bolmont}, E., {Ehrenreich}, D., {et~al.} 2020{\natexlab{a}},
  \aap, 638, A41, \dodoi{10.1051/0004-6361/201937151}

\bibitem[{{Turbet} {et~al.}(2020{\natexlab{b}}){Turbet}, {Gillmann}, {Forget},
  {Baudin}, {Palumbo}, {Head}, \& {Karatekin}}]{turbet2020a}
{Turbet}, M., {Gillmann}, C., {Forget}, F., {et~al.} 2020{\natexlab{b}},
  \icarus, 335, 113419, \dodoi{10.1016/j.icarus.2019.113419}

\bibitem[{{Turbet} {et~al.}(2018){Turbet}, {Bolmont}, {Leconte}, {Forget},
  {Selsis}, {Tobie}, {Caldas}, {Naar}, \& {Gillon}}]{turbet_2018}
{Turbet}, M., {Bolmont}, E., {Leconte}, J., {et~al.} 2018, \aap, 612, A86,
  \dodoi{10.1051/0004-6361/201731620}

\bibitem[{{Unterborn} {et~al.}(2023{\natexlab{a}}){Unterborn}, {Desch},
  {Haldemann}, {Lorenzo}, {Schulze}, {Hinkel}, \& {Panero}}]{unterborn2023}
{Unterborn}, C.~T., {Desch}, S.~J., {Haldemann}, J., {et~al.}
  2023{\natexlab{a}}, \apj, 944, 42, \dodoi{10.3847/1538-4357/acaa3b}

\bibitem[{{Unterborn} {et~al.}(2023{\natexlab{b}}){Unterborn}, {Desch},
  {Haldemann}, {Lorenzo}, {Schulze}, {Hinkel}, \& {Panero}}]{unterborn_2023}
---. 2023{\natexlab{b}}, \apj, 944, 42, \dodoi{10.3847/1538-4357/acaa3b}

\bibitem[{{Unterborn} {et~al.}(2016){Unterborn}, {Dismukes}, \&
  {Panero}}]{unterborn2016}
{Unterborn}, C.~T., {Dismukes}, E.~E., \& {Panero}, W.~R. 2016, \apj, 819, 32,
  \dodoi{10.3847/0004-637X/819/1/32}

\bibitem[{{Vidaurri} {et~al.}(2022){Vidaurri}, {Bastelberger}, {Wolf},
  {Domagal-Goldman}, \& {Kumar Kopparapu}}]{vidaurri2022b}
{Vidaurri}, M.~R., {Bastelberger}, S.~T., {Wolf}, E.~T., {Domagal-Goldman}, S.,
  \& {Kumar Kopparapu}, R. 2022, \psj, 3, 137, \dodoi{10.3847/PSJ/ac68e2}

\bibitem[{{Villanueva} {et~al.}(2018){Villanueva}, {Smith}, {Protopapa},
  {Faggi}, \& {Mandell}}]{Villanueva2018}
{Villanueva}, G.~L., {Smith}, M.~D., {Protopapa}, S., {Faggi}, S., \&
  {Mandell}, A.~M. 2018, \jqsrt, 217, 86, \dodoi{10.1016/j.jqsrt.2018.05.023}

\bibitem[{{Way} \& {Del Genio}(2020)}]{way2020}
{Way}, M.~J., \& {Del Genio}, A.~D. 2020, Journal of Geophysical Research
  (Planets), 125, e06276, \dodoi{10.1029/2019JE006276}

\bibitem[{{Way} {et~al.}(2016){Way}, {Del Genio}, {Kiang}, {Sohl}, {Grinspoon},
  {Aleinov}, {Kelley}, \& {Clune}}]{way2016}
{Way}, M.~J., {Del Genio}, A.~D., {Kiang}, N.~Y., {et~al.} 2016, \grl, 43,
  8376, \dodoi{10.1002/2016GL069790}

\bibitem[{{Way} {et~al.}(2017){Way}, {Aleinov}, {Amundsen}, {Chand ler},
  {Clune}, {Del Genio}, {Fujii}, {Kelley}, {Kiang}, {Sohl}, \&
  {Tsigaridis}}]{way2017b}
{Way}, M.~J., {Aleinov}, I., {Amundsen}, D.~S., {et~al.} 2017, \apjs, 231, 12,
  \dodoi{10.3847/1538-4365/aa7a06}

\bibitem[{{Weiss} \& {Marcy}(2014)}]{weiss2014}
{Weiss}, L.~M., \& {Marcy}, G.~W. 2014, \apjl, 783, L6,
  \dodoi{10.1088/2041-8205/783/1/L6}

\bibitem[{{Widemann} {et~al.}(2023){Widemann}, {Smrekar}, {Garvin},
  {Straume-Lindner}, {Ocampo}, {Schulte}, {Voirin}, {Hensley}, {Dyar},
  {Whitten}, {Nunes}, {Getty}, {Arney}, {Johnson}, {Kohler}, {Spohn},
  {O'Rourke}, {Wilson}, {Way}, {Ostberg}, {Westall}, {H{\"o}ning}, {Jacobson},
  {Salvador}, {Avice}, {Breuer}, {Carter}, {Gilmore}, {Ghail}, {Helbert},
  {Byrne}, {Santos}, {Herrick}, {Izenberg}, {Marcq}, {Rolf}, {Weller},
  {Gillmann}, {Korablev}, {Zelenyi}, {Zasova}, {Gorinov}, {Seth}, {Rao}, \&
  {Desai}}]{widemann2023}
{Widemann}, T., {Smrekar}, S.~E., {Garvin}, J.~B., {et~al.} 2023, \ssr, 219,
  56, \dodoi{10.1007/s11214-023-00992-w}

\bibitem[{{Wogan} {et~al.}(2024){Wogan}, {Batalha}, {Zahnle},
  {Krissansen-Totton}, {Tsai}, \& {Hu}}]{Wogan2024}
{Wogan}, N.~F., {Batalha}, N.~E., {Zahnle}, K.~J., {et~al.} 2024, \apjl, 963,
  L7, \dodoi{10.3847/2041-8213/ad2616}

\bibitem[{{Wolf} {et~al.}(2022){Wolf}, {Kopparapu}, {Haqq-Misra}, \&
  {Fauchez}}]{2022PSJ.....3....7W}
{Wolf}, E.~T., {Kopparapu}, R., {Haqq-Misra}, J., \& {Fauchez}, T.~J. 2022,
  \psj, 3, 7, \dodoi{10.3847/PSJ/ac3f3d}

\bibitem[{{Wolfgang} {et~al.}(2016){Wolfgang}, {Rogers}, \&
  {Ford}}]{wolfgang2016}
{Wolfgang}, A., {Rogers}, L.~A., \& {Ford}, E.~B. 2016, \apj, 825, 19,
  \dodoi{10.3847/0004-637X/825/1/19}

\bibitem[{Wood {et~al.}(2019)Wood, Smythe, \& Harrison}]{Wood2019}
Wood, B.~J., Smythe, D.~J., \& Harrison, T. 2019, Am. Mineral., 104, 844

\bibitem[{{Yang} {et~al.}(2013){Yang}, {Cowan}, \&
  {Abbot}}]{2013ApJ...771L..45Y}
{Yang}, J., {Cowan}, N.~B., \& {Abbot}, D.~S. 2013, \apjl, 771, L45,
  \dodoi{10.1088/2041-8205/771/2/L45}

\bibitem[{{Zhang} \& {Yang}(2020)}]{2020ApJ...901L..36Z}
{Zhang}, Y., \& {Yang}, J. 2020, \apjl, 901, L36,
  \dodoi{10.3847/2041-8213/abb87f}

\end{thebibliography}


\end{document}